\documentclass{emulateapj}
\usepackage{morefloats}
\usepackage{color}
\usepackage{amsmath}
\usepackage{multirow}
\usepackage{graphicx}

\shorttitle{Dynamical masses from global profiles}
\shortauthors{de Blok \& Walter}
\submitted{Accepted for the Astronomical Journal}
\begin{document}
\title{The impact of the gas distribution on the determination of dynamical masses of galaxies using unresolved observations} 

\author{W.J.G. de Blok\altaffilmark{1,2,3} and Fabian Walter\altaffilmark{4}}
\altaffiltext{1}{Netherlands Institute for Radio Astronomy (ASTRON), Postbus 2, 7990 AA Dwingeloo, the Netherlands}
\altaffiltext{2}{Astrophysics, Cosmology and Gravity Centre,
  Univ.\ of Cape Town, Private Bag X3, Rondebosch 7701, South
  Africa}
\altaffiltext{3}{Kapteyn Astronomical Institute, University of
  Groningen, PO Box 800, 9700 AV Groningen, The
  Netherlands}
%\email{blok@astron.nl}
\altaffiltext{4}{Max-Planck Institut f\"ur Astronomie, K\"onigstuhl 17, 69117, Heidelberg, Germany}
%\email{walter@mpia.de}

\begin{abstract}
The dynamical mass ($M_{\rm dyn}$) is a key property of any galaxy,
yet a determination of $M_{\rm dyn}$ is not straight-forward if
spatially resolved measurements are not available. This situation occurs in
single-dish H\,{\sc i} observations of the local universe, but also 
frequently in high-redshift observations. $M_{\rm dyn}$-measurements
in high-redshift galaxies are commonly obtained through observations
of the CO line, the most abundant tracer of the molecular medium. Even
though the CO linewidth can in most cases be determined with
reasonable accuracy, a measurement of the size of the emitting region
is typically challenging given current facilities. We show how the
integrated spectra (`global profiles') of a variety of galaxy models
depend on the spatial distribution of the tracer gas as well as its
velocity dispersion. We demonstrate that the choice of tracer emission
line (e.g., H\,{\sc i}{} tracing extended, `flat', emission
\emph{vs.}\ CO tracing more compact, `exponential', emission)
significantly affects the shape of the global profiles. In particular,
in the case of high ($\sim 50\,$ km\,s$^{-1}$) velocity dispersions,
compact tracers (such as CO) result in Gaussian-like
(non-double-horned) profiles, as is indeed frequently seen in
high-redshift observations.  This leads to significantly different
determinations of $M_{\rm dyn}$ if different distributions of the
tracer material (`flat' vs. `exponential') are considered. We
determine at which radii the rotation curve reaches the rotation
velocity corresponding to the velocity width, and find that for each
tracer this happens at a well-defined radius: H\,{\sc i}{} velocity
widths typically originate at $\sim 5$ optical scale lengths, while CO
velocity widths trace the rotation velocity at $\sim 2$ scale
lengths. We additionally explore other distributions to take into
account that CO distributions at high redshift likely differ from
those at low redshift. Our models, while not trying to reproduce
individual galaxies, define characteristic radii that can be used in
conjunction with the measured velocity widths in order to define
dynamical masses consistent with the assumed gas distribution.
\end{abstract}


\section{Introduction}

The dynamical mass ($M_{\rm dyn}$) of a galaxy is a fundamental
property of galaxies at low and high redshift. In combination with
measurements of the baryonic mass, it provides information on the
amount of dark matter in galaxies.  In the nearby universe, galaxies
can in principle be well resolved and a rotation curve can be derived
to accurately constrain $M_{\rm dyn}$ as well as derive information on the
distribution and importance of dark matter. This is frequently done by
using interferometric observations to measure the Doppler-shift of the
21--cm emission line of neutral hydrogen (H\,{\sc i}{}).

If a galaxy is not sufficiently resolved, then the derived rotation curve
is no longer a true reflection of the mass distribution. However, even
with spectroscopic data that are spatially unresolved or have a low
spatial resolution, the amplitude of the rotational motions can
still be used to derive an approximate value of the dynamical
mass. Usually this is done by deriving the `global profile' of a
galaxy: this shows the emission or flux of a galaxy as a function of
apparent radial (Doppler) velocity. For local spiral galaxies with flat
rotation curves and extended gas disks, the global profile takes the
shape of the well-known `double-horned profile', caused by the
accumulation of Doppler-shifted emission at the projected velocities
corresponding to the amplitude of the flat part of the rotation
curve. In the absence of spatially resolved data, the global profile
is usually the only option for estimating the dynamical mass. This
situation occurs for single-dish H\,{\sc i}{} surveys of the local universe
(such as HIPASS or ALFALFA), but also for interferometric observations
of galaxies at high redshifts.

The velocity width $W$ of the global profile gives an indication of
(twice) the amplitude of the inclination-projected rotation of a
galaxy. Widths are usually determined at 20 or 50 percent of the
maximum value of the profile to avoid being affected by noisy regions
in the spectrum, and are denoted as $W_{20}$ or $W_{50}$,
respectively. An indicative dynamical mass can be defined as $M_{\rm
  dyn} = [(W/(2\sin i))^2R]/G$, where $W$ is the velocity
width\footnote{In practice, $W$ also needs to be corrected for
  instrumental effects such as the finite width of velocity
  channels. We will however ignore this in this paper.}, $i$ is the
inclination of the galaxy, $G$ the gravitational constant, and $R$ a
representative radius. For a good estimate of the total dynamical
mass, this radius should ideally be the maximum radius at which the
tracer can still be observed. Some of the most common choices for $R$
are the radius of the H\,{\sc i}{} disk or the optical disk.  In the
rest of this paper we use $W \equiv W_{50}$.

One important aspect that is often under-appreciated is that global
profiles are not only determined by the rotation curve of a galaxy,
but also by the spatial distribution of the tracer gas within that
galaxy. To date, most measurements of global profiles of low-redshift
galaxies are based on H\,{\sc i}{} measurements. Neutral hydrogen has
the advantage that it usually extends much further out than the
optical disk, thus tracing the rotation at large radii, and in many
cases it comfortably reaches the flat part of the rotation curve. It
also has a high area-covering factor and, to first order, constant
surface density. These three properties together ensure that global
profiles based on H\,{\sc i}{} observations usually yield a good
representation of the dynamics of galaxies. This is also the
underlying reason why measurements of global H\,{\sc i}{} profiles
have been successfully used in evaluating dynamical masses and
relations such as the Tully-Fisher (TF) relation \citep{tf77}.

While H\,{\sc i}{}-based global profiles are therefore an important and
convenient tool for low-redshift studies, this is not the case for
higher redshifts. With current instrumentation, H\,{\sc i}{} emission is
difficult to observe beyond the local universe. Only a relatively
small number of galaxies have been observed directly in H\,{\sc i}{} at larger
redshifts out to $z\sim 0.2$ with significant investments of observing time
(e.g., \citealt{verheijen2007}). Even with large future telescopes
such as the SKA, it is unlikely that large numbers of galaxies will be
routinely observed in emission in H\,{\sc i}{} beyond $z\sim 1$ without major
investments in observing time (see, e.g., \citealt{abdalla2010}).

Recent studies of the gas in galaxies at these higher redshifts have mainly
used the CO line, which is the brightest tracer of the cold molecular gas
component (see \citealt{carilliwalter} for an overview). The CO line is
intrinsically much brighter than the H\,{\sc i}{} line. However, an important
difference with H\,{\sc i}{} is that the CO distribution is known to be more
compact, at least in detailed studies of nearby galaxies. Here the CO
distribution rarely extends beyond the optical radius $R_{25}$. Also, in
contrast with the almost constant H\,{\sc i}{} surface density, the CO surface
density decreases exponentially with radius (e.g., \citealt{schruba11,
  leroy09}).

This immediately implies that CO can trace the rotation curve of a
galaxy only out to a fraction of the radius that would in principle be
traceable by H\,{\sc i}{}. That is, for a given rotation curve, the
global profile as measured in H\,{\sc i}{} will look different from
the one measured in CO.

Finally, global profiles are also affected by the velocity dispersion
of the gas.  Low-redshift disk galaxies are very much dominated by
rotation; the typical H\,{\sc i}{} velocity dispersion of $\sim 11$ km s$^{-1}$
\citep{tamburro09, ianja12} is many times smaller than the rotation
velocity, and the same is true of the low-redshift CO velocity
dispersion \citep{moses13, anahi13}.

However, at higher redshifts, where there are more actively
star-forming environments, the gas velocity dispersion is expected (and
observed) to be much higher. Measurements of the ionized gas phase of
main-sequence star-forming galaxies at $z \sim 1$--$3$ indicate
dispersions of 30--100\,km\,s$^{-1}$ (e.g., \citealt{tacconi13} and
references therein). Direct measurements of the molecular gas
dispersion at these redshifts are rare, but indicate dispersion values
of $\sim 50$ km\,s$^{-1}$ \citep{swinbank11} to $\sim 100$
km\,s$^{-1}$ \citep{hodge12}.  
%These higher velocity dispersions also
%affect the shapes of the global profiles and the ability to derive
%dynamical masses.

A fundamental assumption in the evaluation of a global profile is that
the galaxy under consideration is rotation-dominated, i.e., the system
is not undergoing a merger, is not accreting massive gas reservoirs
that increase the dispersion, and is sufficiently massive that its
kinematics are not entirely dominated by dispersion. At low redshift,
most galaxies are rotation-dominated, and interacting systems are
easily identified by their morphology. At high redshift, the situation
is more difficult to assess from an observational
perspective. However, there is growing evidence that many of the
$L^\star$ galaxies observed at higher redshifts are in fact
rotation-dominated, even if based on their infrared luminosities one
would classify them as ULIRGs (i.e., $L_{\rm FIR}>10^{13}\ L_\odot$;
at low redshift ULIRGS are known to be interacting systems, but they
constitute ``main sequence'' galaxies at high redshift).

There have been numerous attempts to quantify the kinematics of
high-redshift galaxies, mostly using optical emission lines and
integral field units on 10 m-class telescopes. Early studies showed
that $\sim 1/3$ of spatially resolved galaxies were
dispersion-dominated (e.g., \citealt{weiner06, fschreiber09, law09,
  epinat12}). New observations of a sample of $z\sim 2.2$ galaxies
with improved spatial resolution (using AO-assisted imaging;
\citealt{newman13}) have now shown that some of the galaxies thought
to be dispersion-dominated in earlier studies are in fact
rotating. These authors argue that the larger the size of the galaxy,
the more rotation-dominated it is, as is expected (and found in the
local universe) if galaxies have maximum rotation velocities that
scale with size and have velocity dispersions that are independent of
size.  \citet{newman13} also highlight the issue of ``beam smearing'' which
artificially increases the observed dispersion in semi-resolved
measurements.

In summary, even though it is expected that a fraction of
high-redshift galaxies is not rotation dominated (due to mergers
etc.), recent observational evidence clearly demonstrates that a
significant fraction of this galaxy population is rotating. In this
paper we therefore assume galaxies that are rotation dominated, down
to masses where the dispersion becomes a significant contributor.  
%The
%results discussed here therefore do not apply to, e.g., merging or
%interacting galaxies. 
We quantify how the choice of gas tracer and its
intrinsic distribution (from constant surface density to ultra-compact
distributions) changes the estimated dynamical mass of a given galaxy
using a number of galaxy rotation curve templates.
%It is not our
%intention to model individual galaxies in detail. There are still too
%many unknowns in our understanding of galaxy evolution for
%that. Rather, we present an ensemble of models, and leave it to the
%reader to make an informed choice of the appropriate model. 
%The main assumption is that the galaxies are virialised and dominated
%by rotation, and 

In Sect.\ 2 we describe the method we use to derive the profiles and
the assumptions we make in that process.  In Sect.\ 3, we discuss the
resulting velocity widths and their impact on galaxy scaling
relations. Section 4 deals with the problem of finding a length scale
over which dynamical masses are computed.  We summarize our
conclusions in Sect.\ 5.

\section{Generating global profiles}\label{sec:method} 

Here we briefly describe the  method used to generate the
global profiles.  In Sect.~2.1, we first summarize the procedure used
to calculate global profiles given an input rotation curve and a
radial gas surface density distribution. Section~2.2 describes the
properties of the sample of rotation curves we use in this
analysis. Section~2.3 addresses our choices for the radial surface
density profiles.

\subsection{Profiles}

We follow the method described in \citet{obr09} and briefly summarize
the main features here.  The line profile of an edge-on,
optically thin, filled, flat ring with a constant circular velocity
$V_c$ and a luminosity of unity can be derived as follows. The
apparent observed (radial) velocity $V_{\rm obs}$ of a point on the
ring is given by $V_{\rm obs} = V_c \sin \gamma$, where $\gamma$ is
the angle with the line of sight. Evaluating the change in $V_{\rm
  obs}$ as a function of $\gamma$, the normalized edge-on line profile
of the ring can be written (following the notation from \citealt{obr09}) as
\begin{equation}
\tilde{\psi}(V_{\rm obs},V_c) = 
\begin{cases} \frac{1}{\pi\sqrt{V_c^2 - V_{\rm obs}^2}} & \text{if $|V_{\rm obs}| < V_c$}
\\
0 & \text{otherwise.}
\end{cases}
\end{equation}
This equation does not take into account the velocity dispersion of
the gas. A velocity dispersion $\sigma$ can be incorporated in the
above description as a convolution of $\tilde{\psi}$ with a Gaussian
with dispersion $\sigma$, yielding a line profile $\psi$. This
procedure also softens the discontinuities $\tilde{\psi}$ has for
$|V_{\rm obs}| \rightarrow V_c$. We assume the velocity dispersion to
be isotropic.

\citet{obr09} take into account flux normalizations of the profiles.
As we are only concerned with the relative changes of the line
profiles, and linewidths are not affected by normalizations, we here
ignore these normalization factors.

So far, we have assumed rings with a luminosity of unity.  The
distribution of the gas in a galaxy can be taken into account by
multiplying $\psi$ of each ring with the luminosity (or gas mass) in
that ring.  The total global profile $\psi(V_{\rm obs})$ can be
calculated as
\begin{equation} \psi(V_{\rm obs}) = 2 \pi
\sum_{r=0}^{R} r\,\Delta r\, \Sigma(r)\, \psi(V_{\rm obs}, V_c(r)),
\end{equation} where $\Delta r$ is the width of a ring (with $\Delta
r$ chosen to be small compared to the extent of the galaxy),
$\Sigma(r)$ is the azimuthally averaged radial profile of the gas
distribution, $V_c(r)$ the circular velocity curve, which is usually
taken to be equal the rotation curve, and $R$ is the maximum measured
radius of the gas distribution.  Note that we do not take into account
optical depth effects.

The line profile of a non-edge-on galaxy can be calculated by
multiplying $\tilde{\psi}$ in Eqn.\ (1) with $\sin i$ before
convolving with the velocity dispersion Gaussian.  The procedure
described here thus allows calculation of global profiles for any
given combination of surface density profile and rotation curve and
arbitrary values of inclination and velocity dispersion.

\subsection{Rotation curves}

To calculate global profiles comparable to those typically observed
(at least at low redshift), we need representative rotation curves as
input, preferably over a comprehensive range of galaxy masses and
luminosities.  We here use a set of template rotation curves from
\citet{cat06}.  They used long-slit optical rotation curves and
$I$-band photometry of $\sim 2200$ low-redshift disk galaxies to
construct average template rotation curves for ten separate luminosity
classes, spanning six magnitudes in $I$-band luminosity, ranging from
$M_I = -23.8$ to $M_I = -19.0$ (approximately equal to a range from
$M_B \sim -22.0$ to $M_B \sim -17.2$, assuming $B-I \sim 1.8$;
cf.\ \citealt{dejong_cols}, their fig.\ 10 and table 2).  For
convenience we denote these luminosity classes with a numerical code
$t$, with $t=0$ corresponding to the brightest luminosity class and
$t=9$ to the faintest luminosity class (see also
Tab.~\ref{tab:template}).

The template curves as presented in \citet{cat06} are expressed in
terms of the exponential disk scale length $h$ and in all cases extend
out to $\sim 5h$.  \citet{cat06} parameterise these curves using the
so-called polyex model \citep{gh2002},
\begin{equation} V_{PE}(r) =
V_0 \left( 1-e^{-r/r_{PE}}\right) \left(1+\frac{\alpha\, r}{r_{PE}}
\right).  
\end{equation} 

\begin{figure*}
\includegraphics[width=0.45\hsize]{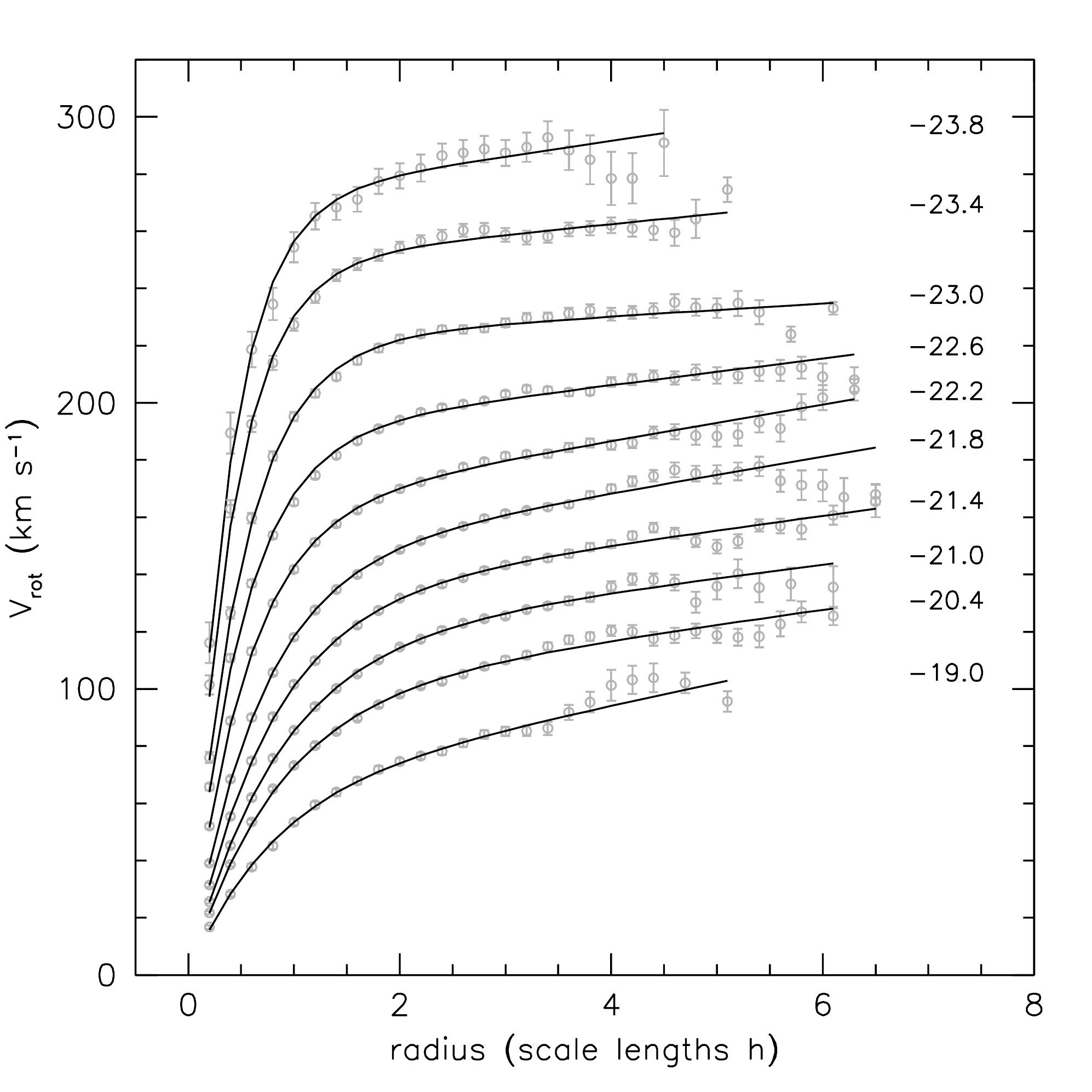}
\includegraphics[width=0.45\hsize]{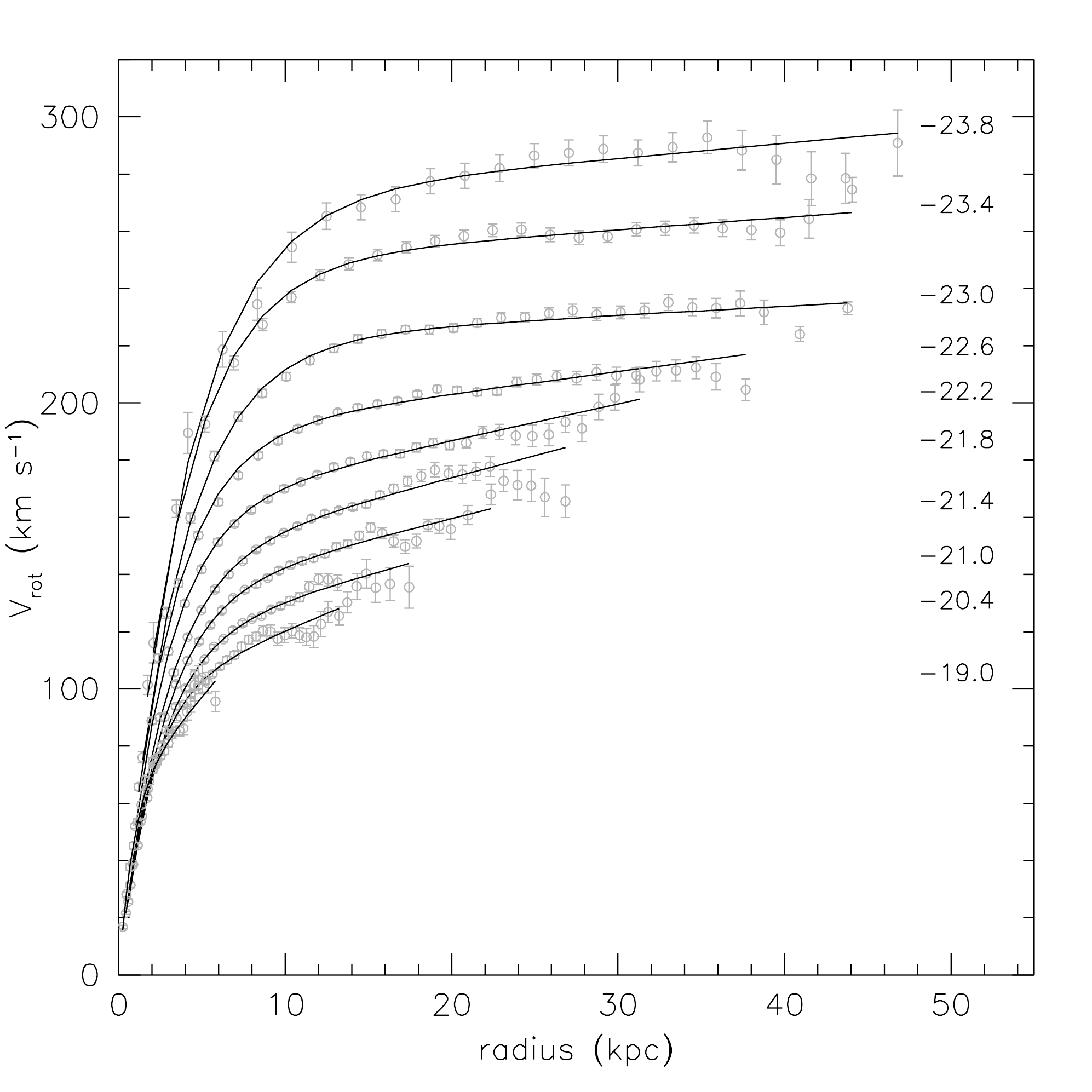}
\caption{Template rotation curves from \citet{cat06} (grey
  symbols). The polyex fits are overplotted as the black curves. The
  average luminosity corresponding to each template curve is given to
  the right of the terminal velocity of each curve. The left-hand
  panel shows the curves plotted with the radius expressed in scale
  lengths $h$. The right-hand panel shows the same curves with the
  radius in kpc.  These are derived by multiplying the radial
  scale in units of scale length by the
  value of $h$ as derived in the text and given in Table 1.
\label{fig:templates}}
\end{figure*}

Here $V_0$, $r_{PE}$ and $\alpha$ describe the amplitude of the
rotation curve, the steepness of the inner rise of the curve, and the
slope of the outer part, respectively.  The average luminosities of
each luminosity class and the parameters of the corresponding polyex
fits, taken from \citet{cat06}, are listed in
Tab.~\ref{tab:template}.  For a full discussion of curves and fits we
refer to \citet{cat06}. In the following we will be adopting the
polyex fits as our fiducial template curves. This has the advantage that
small-scale variations in the rotation curve do not affect the
results. The curves and polyex fits with radii expressed in scale
lengths are shown in the left panel in Fig.~\ref{fig:templates}.

We will be varying the distribution of the gas tracer in our models
and therefore need to define a representative reference length scale
for each template rotation curve.  We here choose the optical
exponential disk scale length $h$, which, as noted, is also the unit
of radius of the \citet{cat06} template curves.

It is useful to also express the rotation curve radii in kpc
units, and we calculate values for the scale lengths $h$ as follows.
First, we assume that the template galaxies are Freeman-disks (i.e.,
that they have a central disk surface brightness of $\mu_0^B = 21.65$
mag arcsec$^{-2}$), which is a reasonable assumption for this range in
luminosity (see, e.g., \citealt{dejong_mu}). Note that as we are
ignoring normalisation factors in this analysis, the exact choice of
central surface brightness value is not critical.  Using an average
colour for these galaxies of $B-I \sim 1.8$ (as also used above), we
calculate the Freeman-value in the $I$-band. By assuming exponential
disks, we can use $L_I = 2\pi\Sigma_0^I h^2$ to calculate the
exponential scale length $h$. Here $L_I$ and $\Sigma_0^I$ are the
$I$-band luminosity and Freeman central surface brightness value,
expressed in $L_{\odot}$ and $L_{\odot}\, {\rm pc}^{-2}$,
respectively. The scale lengths $h$ that are derived for each template
rotation curve are listed in Tab.~\ref{tab:template}. The right panel
in Fig.~\ref{fig:templates} shows the resulting template curves and
polyex fits with the radii in kpc scaled using the appropriate scale
lengths.  The left panel in Fig.~\ref{fig:templates}  shows
that all curves extend to $\sim 5h$; the right panel shows that the
decrease in scale length from high to low luminosity (or,
equivalently, galaxy mass) results in the curves extending less far
(in terms of kpc) as luminosity decreases.

A maximum radius of $5h$ agrees well with the typical sizes of H\,{\sc
  i}{} disks in nearby disk galaxies.  For a sample of 108 nearby
gas-rich disk galaxies, \citet{broeilsrhee} find a relation between
the H\,{\sc i}{} radius $R_{\rm HI}$ (defined at the
$1\,M_{\odot}\,{\rm pc}^{-2}$ column density level) and the optical
radius $R_{25}$ (defined at the 25 $B$-mag arcsec$^{-2}$ level) of
$R_{\rm HI} \sim (1.70\pm 0.16)\, R_{25}$. For nearby disk galaxies
$R_{25} \simeq 3.2h$, yielding an H\,{\sc i}{} radius $R_{\rm HI} \sim
(5.4 \pm 0.5)h$. The $5h$ maximum radius used here is therefore a
reasonable approximation of that found in real disk galaxies.

The rotation curves described here are derived from local
galaxies. Are they also applicable to high-redshift galaxies?  As
described in the Introductions, there is good evidence that a
substantial fraction of galaxies at higher redshifts are domination by
rotation. In addition, the TF relation, one of the most well-defined
relations involving rotating disk galaxies, seems to be in place
already by $z\sim 1.7$ (e.g., \citealt{miller}), and there are
indications it may already exist by $z\sim 3$ (e.g.,
\citealt{gnerucci}), though with an offset in stellar mass. Galaxies
with well-defined rotation curves with a range in maximum rotation
velocities overlapping with that observed locally can thus be presumed
to exist at higher redshifts as well. We will later show that the
global profiles are only moderately sensitive to the exact shape of
the rotation curve.

Whether the gas distributions at higher redshifts are similar to those
found locally is an open question. However, as described in the
following subsection, we circumvent this by calculating global
profiles for a large range of surface density profiles.

\begin{deluxetable*}{ccccccrc}
\tabletypesize{\scriptsize}
\tablewidth{0pt}
\tablecaption{Properties of template
  rotation curves.\label{tab:template}}
\tablehead{\colhead{$t$} & \colhead{$M_I$} &  \colhead{$\Delta M_I$} &
   \colhead{$V_0$} &  \colhead{$r_{PE}/h$} &  \colhead{$\alpha$} &
   \colhead{$h$} &  \colhead{$V_{\rm max}$}\\  \colhead{} &
   \colhead{(mag)} &  \colhead{(mag)} &  \colhead{(km\,s$^{-1}$)} &
   \colhead{} &  \colhead{} &  \colhead{(kpc)} &
   \colhead{(km\,s$^{-1}$)}\\  \colhead{(1)} &  \colhead{(2)} &
   \colhead{(3)} &  \colhead{(4)} &  \colhead{(5)} &  \colhead{(6)} &
   \colhead{(7)} &  \colhead{(8)}}
\startdata
0 & $-23.8$ & 0.4 & 270 &
0.37 & 0.007 & 10.4 & 294 \\ 1 & $-23.4$ & 0.4 & 248 & 0.40 & 0.006 &
8.6 & 267 \\ 2 & $-23.0$ & 0.4 & 221 & 0.48 & 0.005 & 7.2 & 235 \\ 3 &
$-22.6$ & 0.4 & 188 & 0.48 & 0.012 & 6.0 & 217 \\ 4 & $-22.2$ & 0.4 &
161 & 0.52 & 0.021 & 5.0 & 201 \\ 5 & $-21.8$ & 0.4 & 143 & 0.64 &
0.028 & 4.1 & 184 \\ 6 & $-21.4$ & 0.4 & 131 & 0.73 & 0.028 & 3.4 &
163 \\ 7 & $-21.0$ & 0.4 & 116 & 0.81 & 0.033 & 2.9 & 144 \\ 8 &
$-20.4$ & 0.8 & 97 & 0.80 & 0.042 & 2.1 & 128\\ 9 & $-19.0$ & 2.0 & 64
& 0.72 & 0.087 & 1.1 & 102 \\
\enddata
\tablecomments{
(1) Luminosity
  $t$-type. (2) Average luminosity of galaxies in this bin. (3) Spread
  in luminosity of galaxies in this bin. (4) Velocity amplitude of
  polyex fit (eq.~3). (5) Steepness of inner polyex fit rotation
  curve, expressed in scale lengths. (6) Slope of outer polyex fit
  rotation curve. (7) Exponential disk scale length in kpc. (8)
  Maximum velocity of polyex fit rotation curve. Columns (2)--(6) are
  from \citet{cat06}.}
\end{deluxetable*} 

\subsection{Radial gas distributions}

We adopt the scale lengths $h$ derived above as the `standard' values
appropriate for each luminosity class $t$. That is, if in a template
galaxy the gas tracer is distributed like the stars, it also has a
scale length $h$. In evaluating global profiles, we vary the scale
length of the gas distribution with respect to this reference value
$h$. For each template rotation curve (or luminosity bin $t$) we
assume gas distributions with scale lengths $\ell = h/4$, $h/2$, $h$,
$2h$ and $4h$, as well as a constant (`flat') radial surface density,
the latter corresponding to the situation typically found in H\,{\sc
  i}{} disks in galaxies.

In addition, we evaluate the distribution given in \citet{schruba11}
(hereafter S11) who found that the CO radial profiles in a number of
nearby disk galaxies could be described as an exponential with $\ell
\simeq 0.2 R_{25}$. Again using that $R_{25} \simeq 3.2h$, we can
describe the  empirical S11 distribution using a scale
length $\ell_{\rm S11} = 0.64h$.

In constructing the global profiles, we take into account that the
template rotation curves themselves terminate at $\sim 5h$.  Models
which, due to a large value of $\ell$, give rise to very extended gas
distributions are truncated at $R=5h$, thus taking into account the
typical ``largest radius'' accessible in observations of (nearby)
galaxies.

In Appendix A we compare a number of our analytically derived global
profiles with those derived from more CPU-intensive numerical,
three-dimensional models.  We find very good agreement and proceed to
use the analytical profiles in the rest of this paper.

The large range in radial density profiles presented here enables
evaluation of profiles based on a situation that deviates from what is
found locally. If, for example, the observed CO distribution at high
redshift is more compact than what is observed at low redshift
(because high-$J$ CO transitions have been observed, or because of a
radially changing CO-to-H$_2$ conversion factor, e.g., due to steep
metallicity gradients), then it is easy to calculate the profile for a
smaller $\ell$-value. We emphasize here that we are not attempting to
provide a definitive description of a high-redshift galaxy. Rather, we
evaluate profiles covering a range in maximum rotation velocities and
gas densities that overlaps with what is found observationally.  A
comparison with these observations should therefore be made under the
following assumptions: 1) galaxies are virialised and dominated by
rotation; 2) galaxies do not suffer from major interactions and/or
asymmetries (though the latter will have only a minor impact on
$W_{50}$); 3) rotation curves are to first order independent of
redshift (as hinted at by the existence of a high-redshift TF
relation); 4) the range in gas density profiles presented here covers
that found in real galaxies.

Also note that the surface density of the tracer component is
not relevant for our models, as we only evaluate normalized global
profiles. In order to model more compact or more extended gas
distributions only the scale length $\ell$ needs to be adjusted.

\section{Results} 

\subsection{Comparing velocity widths} 

For each of the template rotation curves ($t$-types), we calculate
global profiles, assuming exponential radial gas density profiles with
scale lengths $\ell = (h/4, h/2, h, 2h, 4h)$, where $h$ is the
`standard' scale length for each template galaxy class as listed in
Tab.~1.  We also calculate global profiles for a constant
surface density filled disk and an exponential disk with an S11 scale
length $\ell_{\rm S11} = 0.64h$.  Gas distributions are in all cases
truncated at a radius of $5h$.

We calculated many sets of global profiles assuming different
combinations of inclination and velocity dispersion. We here
concentrate on the profiles of edge-on galaxies with $\sigma = 10$ and
$50$ km s$^{-1}$, representative of velocity dispersions typically
found in low- and high-redshift galaxies, respectively.  Appendix B
presents profiles at inclinations of 60$^{\circ}$ and 30$^{\circ}$.

\begin{figure*}
\includegraphics[width=\hsize]{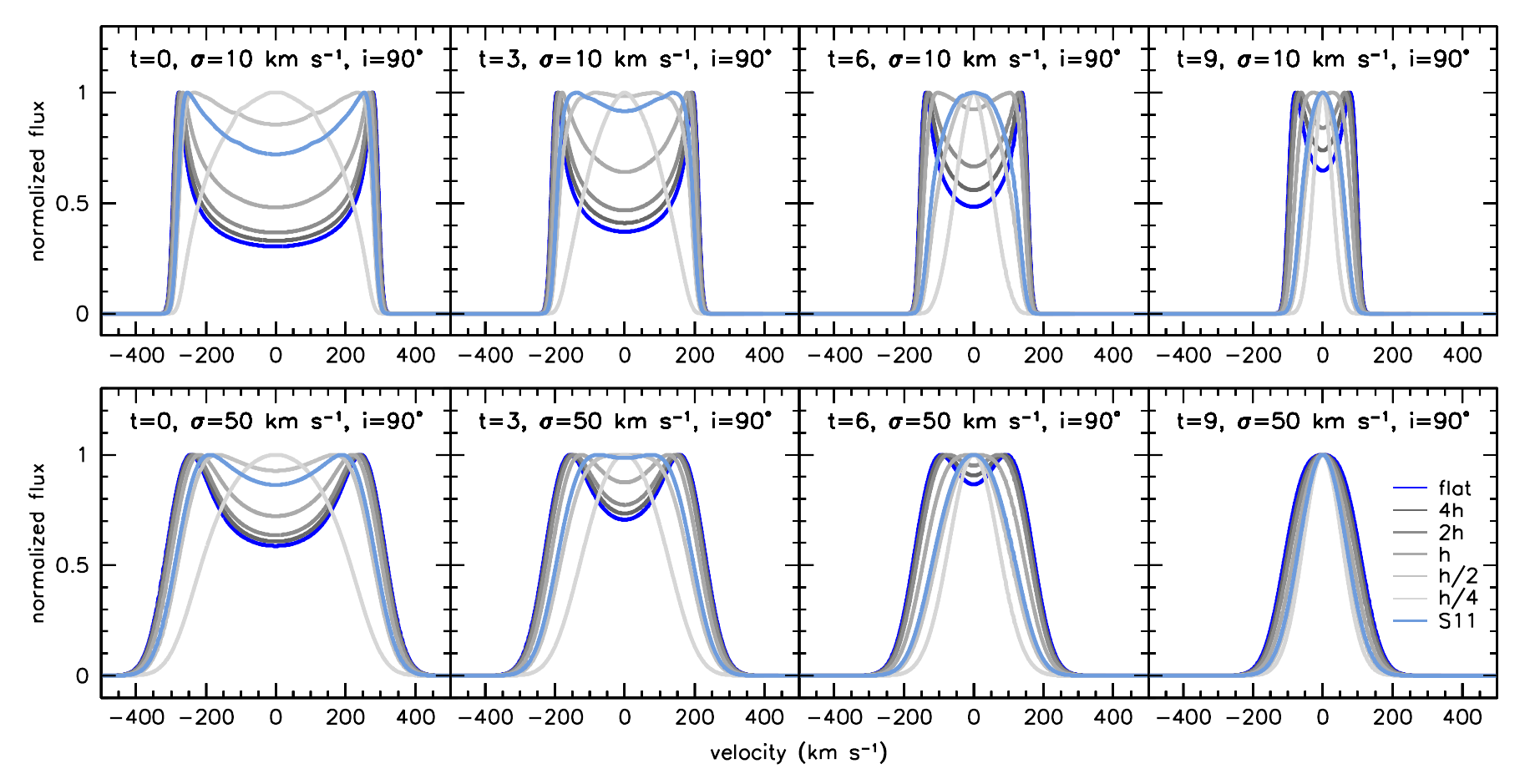}
\caption{Global profiles of edge-on galaxies for a number of
  luminosity classes ($t=0,3,6,9$). The top row shows profiles with
  velocity dispersion $\sigma = 10$ km s$^{-1}$, the bottom row
  assumes $\sigma = 50$ km s$^{-1}$. Different profile colors indicate
  different steepnesses of the tracer gas profile, as indicated in the
  bottom--right panel, where the value of $\ell$ is indicated. The
  profiles derived assuming a constant surface density, filled disk
  are shown in dark blue (indicated with `flat' in the bottom-right
  panel) and correspond to what is typically found in H\,{\sc i}{}
  observations of nearby disk galaxies. The light-blue profiles (`S11'
  in the bottom-right panel) assume an exponential disk with scale
  length $\ell = 0.64h$, as found by \citet{schruba11} for nearby
  galaxies.\label{fig:profiles_i90}}
\end{figure*}

Fig.~\ref{fig:profiles_i90} shows the profiles for $t=(0,3,6,9)$
edge-on galaxies, assuming $\sigma = 10$ km\,s$^{-1}$ (top row) and
$50$ km\,s$^{-1}$ (bottom row) velocity dispersion and using the range
of $\ell$-definitions given above. The steepness of the gas profiles
is indicated by the color of the profile (as indicated in the
bottom-right panel). 

In Fig.~\ref{fig:profiles_i90}, we see that for the low-$t$,
low-dispersion, constant-surface-density models the double-horned
signature of the profiles is very pronounced. In contrast, for very steep gas
profiles, where the density has already become negligible at radii
where the rotation velocity has not reached the flat part yet, the
horns disappear, resulting in a more Gaussian profile. This is an
effect that we see for all parameter combinations. It is most
pronounced at higher $t$-values, where due to the slower rising
rotation curves, the horns already disappear for moderately steep gas
profiles.

As the value of the velocity dispersion increases (bottom row of
Fig.~\ref{fig:profiles_i90}), we see that the horns are less
pronounced, with the profiles becoming Gaussian or flat-topped over a
large range of parameter combinations. This could be an explanation
why many CO spectra of galaxies at high redshift do not exhibit the
double-horn shape. For example, the majority of the quasar host
galaxies show profiles that are well fit by just a simple Gaussian
(e.g., \citealt{riechers11}). This is also the case for most of the
sub-millimeter galaxies (e.g., \citealt{bothwell13}). Recent large
samples of main sequence galaxies have also revealed predominantly
Gaussian CO profile shapes (e.g., \citealt{tacconi13}).  The
double-horned profiles that are observed at high redshifts are most
often associated with massive galaxies, which must have steeply rising
rotation curves or very extended molecular gas disks to ensure that
the double-horned signature remains visible.  For example, the
double-horned CO profiles shown in \citet{daddi10} belong to
high-redshift galaxies with velocity widths 400--600 km s$^{-1}$,
which, in terms of velocity width, are comparable with the $t=0$ or
$t=1$ galaxies as discussed here (cf.\ Fig.~\ref{fig:profiles_i90}).

\begin{figure*}
\includegraphics[width=0.9\hsize]{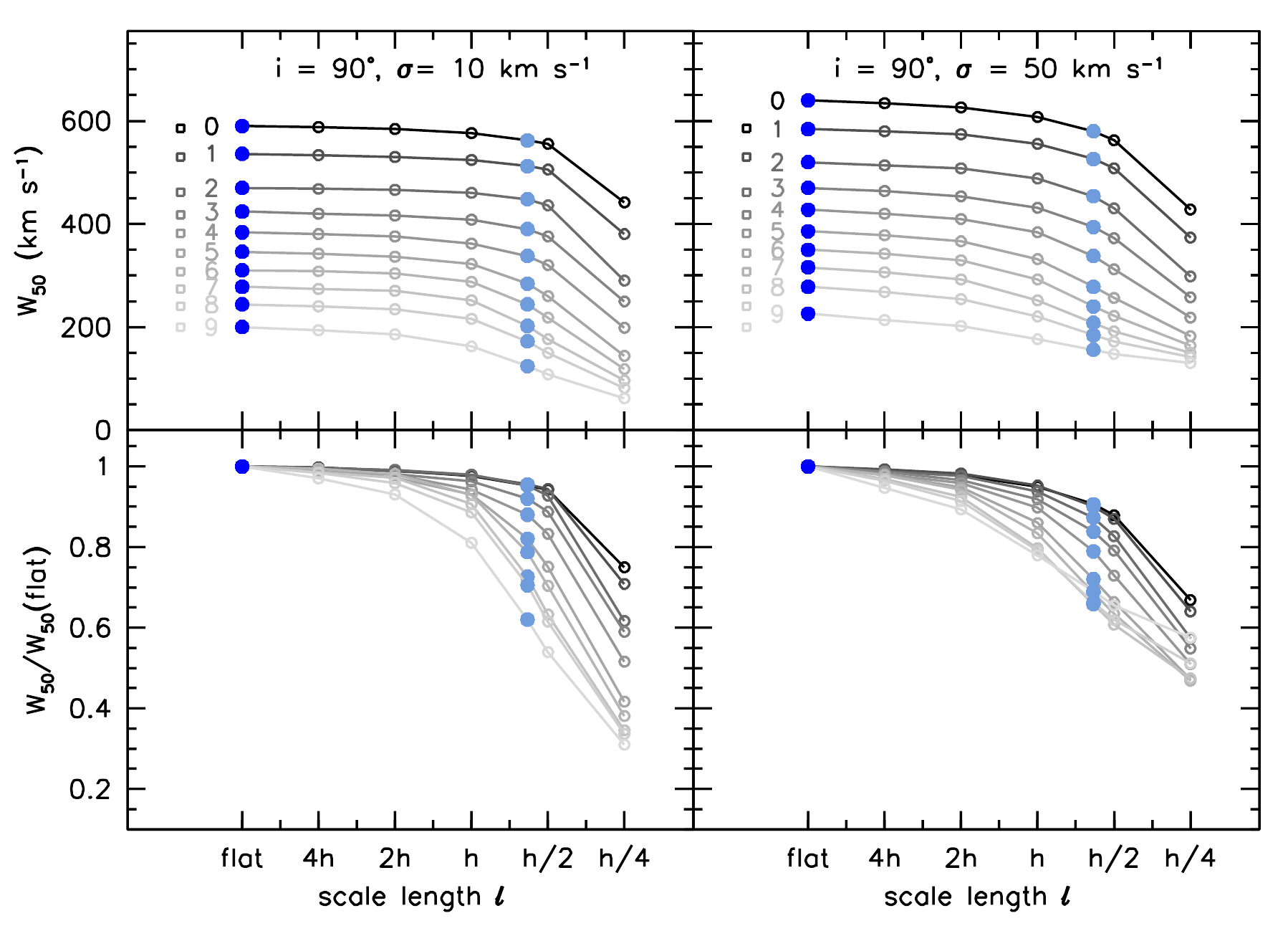}
\caption{$W_{50}$ values for velocity dispersions of 10 km s$^{-1}$
  (left) and 50 km s$^{-1}$ (right) as a function of profile steepness
  (scale length $\ell$) for edge-on galaxies. Grey-scales represent the
  $t$ type (also indicated in the top panels). Dark-blue indicates
  flat density profiles (`H\,{\sc i}{}'), light-blue indicates S11 profiles
  (`CO'). For comparison, the small squares in the top panels indicate
  the values for the flat density profiles, but assuming a zero velocity
  dispersion.\label{fig:w50}}
\end{figure*}

In the majority of cases, the apparent velocity width of the profile
becomes discernibly smaller as the steepness of the gas surface
density profile increases. The change can be a significant fraction of
the velocity width, especially for the lower luminosity galaxies. This
is quantified in Fig.~\ref{fig:w50}, where we show how $W_{50}$
changes as a function of the gas distribution\footnote{For most
  profiles this width is unambiguously defined. In cases where the 50
  percent level crosses the horns, we define $W_{50}$ to be the
  difference between the two outermost crossings (i.e., the ones with
  the highest and lowest velocities).} (shown here for only for
edge-on galaxies; other
inclinations are shown in Appendix B).

The values of $W_{50}$ are largely unaffected for gas distributions
that have scale lengths $\ell > h$ (i.e., extended distributions).
For smaller values of $\ell$ (compact distributions), the measured
values decrease rapidly, especially for the less luminous galaxies
(high $t$-values). For the S11 profile, the ratio of the measured
$W_{50}$ value and the value for a constant surface density disk
decreases from slightly less than unity for $t=0$ to $\sim 0.6$ for
$t=9$. These trends are visible for both values of the velocity
dispersion.  Note that for $\sigma=50$ km s$^{-1}$, we see a different
behavior for the smallest $\ell$ values. In this case the velocity
dispersion dominates the measured $W_{50}$ values, and we see a
flattening of the decrease in $W_{50}$ as a function of decreasing $\ell$.

\subsection{The Tully-Fisher relation}

These findings have direct consequences for the interpretation of
scaling relations involving a velocity width. Perhaps the most famous
of these is the Tully-Fisher (TF) relation \citep{tf77}. This relation
between the luminosity and velocity linewidth of galaxies has been
used as a distance indicator, but the apparent universality of the TF
relation has meant that it has also become an important part of
studies of the dark and baryonic matter content of galaxies (e.g.,
\citealt{mcgaugh2000,mcgaugh12}).

The TF relation is most often studied using H\,{\sc i}{} as a tracer
for the dynamics.  Here we investigate whether the parameters of the
TF relation, in particular the slope, are sensitive to changes in the
radial profile of the tracer gas. Aspects of this question were
already investigated in papers by \citet{tutui} and \citet{lavezzi}.
\citet{tutui} already note the effect that the different CO
and H\,{\sc i}{} distributions will have.

\begin{figure*}
\includegraphics[width=0.9\hsize]{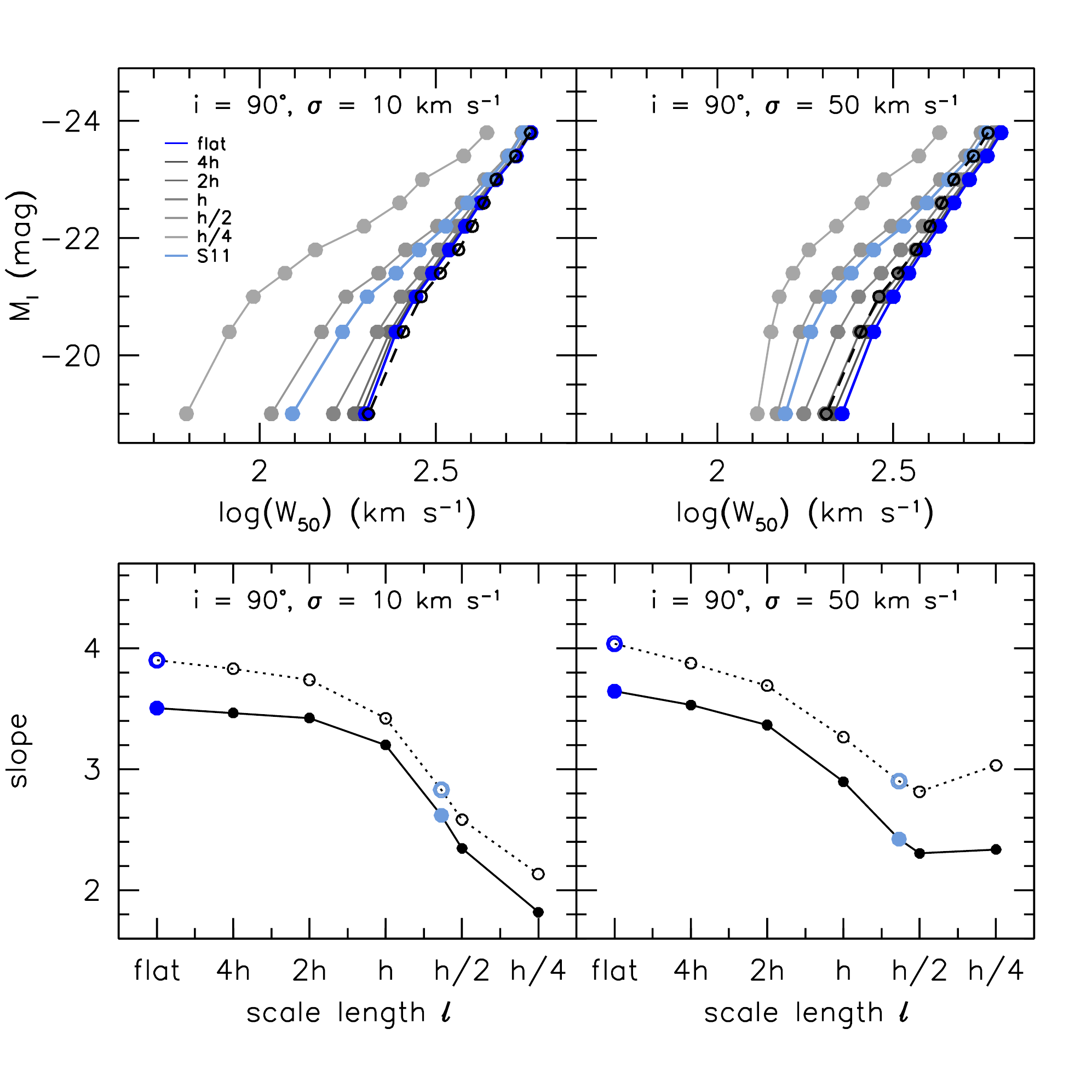}
\caption{{\bf Top-left:} TF relations for various gas tracer profile
  steepnesses assuming $\sigma = 10$ km
  s$^{-1}$. Steepnesses are indicated in the panel.  The black dashed-line
  with open circles represents the TF relation derived using the outer
  velocity in the rotation curves.  {\bf Bottom-left:} fitted slopes of
  the TF relations.  Dotted lines and open symbols show slopes of the
  full TF relations $(t=0-9)$, full lines and filled symbols show
  slopes derived using only points with $M_I<-20.5$ $(t=0-7)$. Two
  measures of the slope are used due to the break in the TF around
  $M_I\sim -21$.  {\bf Top-right:} As top-left, but for $\sigma=50$
  km s$^{-1}$. {\bf Bottom-right:} As top-right, but for $\sigma = 50$
  km s$^{-1}$.\label{fig:tf}}
\end{figure*}

In Fig.\ \ref{fig:tf} (top row), we show the TF relations derived
using our measured velocity widths for the various assumptions on the
steepness of the tracer gas profile and a velocity dispersion $\sigma
= 10$ km s$^{-1}$. We use the absolute $I$-band magnitudes
corresponding to each $t$-type, as given in Tab.~1. For comparison,
the TF relation as defined by the rotation velocities at the outermost
point of the template rotation curves is also shown.  Note that
\citet{verheijen2001} finds an $I$-band TF slope of $3.8 \pm 0.1$,
based on global profile velocity widths in an H\,{\sc i}{}-study of
the TF relation of regular disk galaxies in the Ursa Major cluster
(the value quoted here applies to their `DE sample'). In general,
H\,{\sc i}{}-based studies of the TF relation tend to find slopes
$\approx 4$ (e.g., \citealt{mcgaugh2000,verheijen2001}).

From Fig.\ \ref{fig:tf} it is clear that the slope of the TF relation
decreases as the steepness of the tracer gas profile
increases. Moderately steep density profiles already result in a
significant change in the slope.

Note that the TF relation based on the flat density profile shows a
change in the slope around $M_I \sim -21$. We therefore determine two
values for the slope of each TF relation. The first one is determined
using the full luminosity range, i.e., all $t$-types. The second one
only uses the galaxies with $M_I < -20.5$, i.e., only $t=0-7$ (the
galaxies above the break).  Figure \ref{fig:tf} (right) shows the
resulting TF-slopes as a function of the steepness of the gas
profile. For the flat and extended density profiles we find TF-slopes
that are in reasonable agreement with those found in the detailed TF
study by \citet{verheijen2001} mentioned earlier, but note that the
TF-slope starts to decrease around $\ell = h$. The slope of the TF
relation derived assuming S11 density profiles has decreased by a full
unit compared to the value found for flat density profiles.  The slope
thus depends on the steepness of the density profile, and for the
models considered here a `CO'-TF would be less steep than an `H\,{\sc
  i}{}'-TF. This conclusion applies to any density profile steeper
than $\ell \sim h$.

Results for a velocity dispersion of 50 km s$^{-1}$ are similar
(Fig.~\ref{fig:tf}, bottom-row), with the one difference that at low
linewidths the dynamics are clearly dominated by the
dispersion. Nevertheless, also here, a TF measured in CO would have a
more shallow slope than an H\,{\sc i}-based TF.

\section{Comparing dynamical masses}

When velocity widths are used to estimate dynamical masses, the
differences due to the different tracer gas profiles directly
propagate into the mass estimates through the \emph{square} of the
difference via $M_{\rm dyn} \sim W_{50}^2\times R$.  A second
complication is that a radius $R$ is needed for the calculation of the
dynamical mass. In practice, when one has to resort to using $W_{50}$
for dynamical mass estimates, this usually means that the observations
of the gas component are unresolved or, at best, semi-resolved, and
$R$ cannot be measured directly.  It is therefore customary to
\emph{assume} a radius that is deemed appropriate for a given
tracer. In the case of high-redshift CO observations of main sequence
galaxies, this is typically chosen to be the optical radius, based on
the results found for nearby galaxies (e.g., \citealt{leroy08,
  leroy13}) and the few marginally resolved CO measurements at high
redshift (e.g., \citealt{daddi10, tacconi13}). In dealing with these
observations, one has to be aware of possible mismatches between the
distribution of the tracer and the assumed radius for the dynamical
mass.

\begin{figure*}
\includegraphics[width=\hsize]{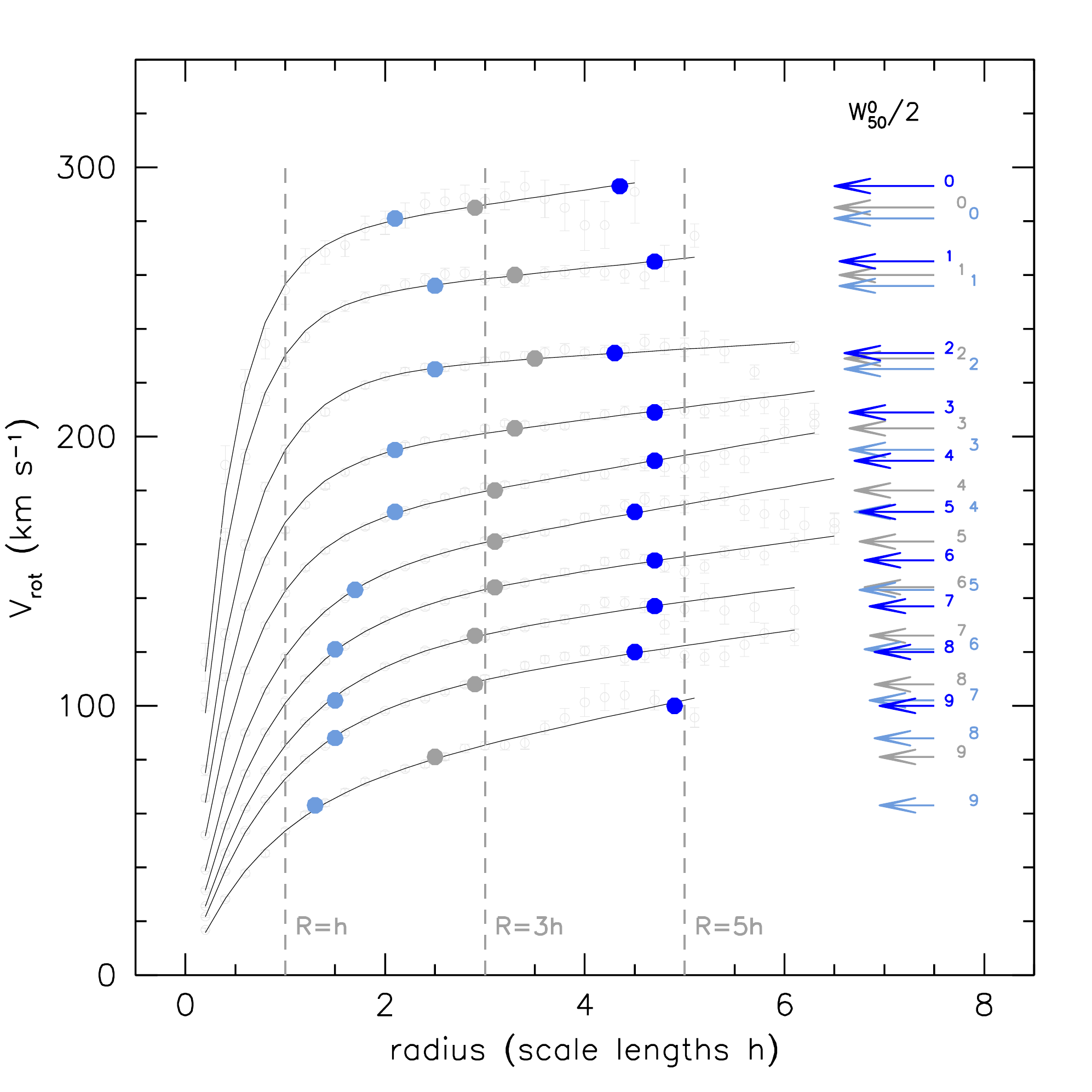}
\caption{Template rotation curves with radii expressed in scale
  lengths. See Fig.~1 for luminosities and $t$-type numbering. The
  arrows indicate the $W_{50}^0/2$ values assuming zero dispersion for
  the flat (blue), exponential (gray) and S11 (light-blue) surface
  density profiles for each of the $t$-types as indicated to the right
  of the arrows. The filled coloured circles overplotted on top of the
  curves indicate the radii where the respective $W^0_{50}/2$ values
  occur, using the same colours as the corresponding
  arrows.\label{fig:curvescale}}
\end{figure*}

\subsection{Tests using rotation curves}

To address the problem of choosing the appropriate radius, we will
here use our models of the global profiles to identify radii that are
intrinsic to the tracer distribution used.  In combination with the
global profile, this radius can then be used to give a consistent estimate
  of the dynamical mass assuming a radius appropriate for the chosen
  gas distribution.

In the following we will, for brevity, only discuss the cases for flat
(`H\,{\sc i}{}'), $\ell=h$ (`exponential') and $\ell = \ell_{\rm S11}$
(`CO') density profiles,  but results for other density profiles can be derived in a similar manner.  For each of the three surface density
distributions, we identify in the template rotation curves those radii
that show the same rotation velocity as the $W_{50}/2$ values of the
corresponding global profile.  As the rotation curve velocities do not
contain a dispersion component, we compare with $W_{50}$ values
derived from global profiles we calculate assuming zero dispersion,
hereafter denoted as $W_{50}^0$. We return to the impact of the
velocity dispersion at the end of this subsection.

Figure \ref{fig:curvescale} shows the template rotation curves again,
with the radii given in units of $h$. In the Figure, we also indicate
with arrows the measured  $W_{50}^0/2$ values for each
combination of $t$-type and the three density profiles (flat,
exponential and S11). We also indicate the radii at which these
velocities occur. It is immediately clear that each type of density
profile is associated with its own distinct range in radii.  For the
flat density profiles, the $W_{50}^0$ value measured is associated with
a radius of $(4.6 \pm 0.2)h$; the exponential profile with a radius of
$(3.1 \pm 0.3)h$, and the S11 profile with $(1.9 \pm 0.4)h$.

A $W_{50}^0$ velocity width determined in H\,{\sc i}{} (i.e., with a
flat density profile) is thus measuring the rotation velocity that is
typically found at $\sim 5h$, and the latter is therefore the
appropriate radius to use in calculating the dynamical mass. 
  Assuming an S11 profile instead (corresponding to a low-redshift CO
  measurement), will yield a different velocity width due to the
steeper density profile; here the appropriate radius is $\sim 2h$.
Under the assumptions made here, velocity widths measured in H\,{\sc
  i}{} are thus representative of the dynamical mass enclosed within a
$\sim 5h$ radius, while CO measurements trace the dynamical mass
enclosed within a $\sim 2h$ radius. The CO distribution and radii
  discussed here apply, strictly speaking, only to local
  galaxies. Should the CO distribution in high-redshift galaxies be
  different, then an appropriate, different value of $\ell$ can be
  chosen and corresponding radii derived.

\begin{figure}
\includegraphics[width=\hsize]{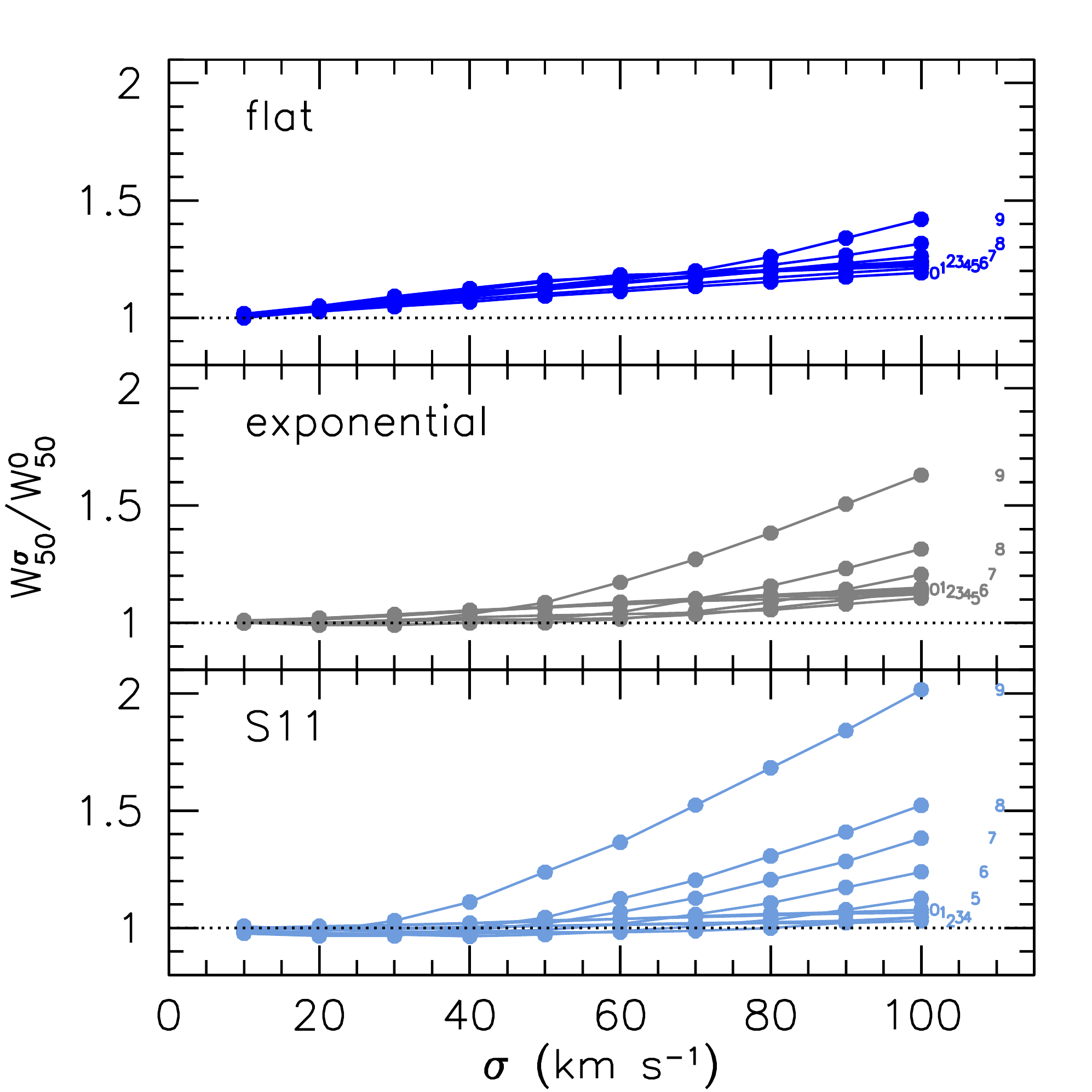}
\caption{Ratio of the velocity widths that include velocity dispersion
  (as indicated on the horizontal axis) and the zero-dispersion
  velocity widths. From top to bottom are shown the ratios for the
  flat, exponential and S11 density profiles. In each panel curves are shown for
  each $t$-type, as indicated by the numbers in the right-most part of
  the plot. Corrections more than $\sim 15-20\%$ should be treated with care, as they indicate the global profile is not dominated by rotation. 
\label{fig:dispratio}}
\end{figure}

%%%%%%%%%%%%%%%%%%

Dynamics of real galaxies contain a dispersion term: for the three
density profiles we discuss here we have calculated the ratio of
velocity widths that include a dispersion and the zero-dispersion
velocity widths. These ratios are plotted in Fig.~\ref{fig:dispratio}.
For the flat density profile, corrections rise approximately linearly
with dispersion, until they reach values of $\sim 20\%$ at $\sigma =
100$ km s$^{-1}$ for all but the latest $t$-types where the
corrections are much higher. The exponential density profiles shows
somewhat smaller corrections than the flat density profile, peaking at
$\sim 15\%$, with again the late $t$-types showing larger
corrections. The S11 profile yields larger corrections overall, with
the latest types already diverging at intermediate dispersions. Note
that for the S11 profiles, and to a lesser extent also the exponential
profiles, we find some ratios that are slightly less than unity. This
is a real effect, with the larger dispersions moving flux to more
extreme velocities, giving the central profiles a slightly more narrow
core.

The larger corrections with increasing density profile
steepness show how the steep density profiles ensure that a relatively
small part of the inner rotation curve is sampled, causing  the
velocity dispersion to have a higher impact.  Profiles with corrections
that deviate strongly from the locus of curves shown in
Fig.\ \ref{fig:dispratio} are thus unlikely to retain much information
about the rotation of the galaxy, as can also be deduced from comparing
profile shapes in Fig.\ \ref{fig:profiles_i90}. Corrections higher
than $\sim 15\%$ to $\sim 20\%$ should therefore be treated with caution
as the profile retains little rotational information.

Given an observed global profile and assumed mass density profile, one
can therefore estimate a representative dynamical mass using the
velocity-dispersion-corrected velocity width and the corresponding
characteristic radius.

\subsection{Tests using global profiles}

In the previous subsection we derived the characteristic radii using
the input rotation curves. This is an illustrative way of defining and
locating the characteristic radii, but we can also derive these
directly from the global profile. This has the advantage that the
velocity dispersion can be taken into account directly. To derive
characteristic radii from the profile, recall Eqn.\ (2). This
describes the value of the global profile at velocity $V_{\rm obs}$ as
the weighted sum of the fluxes that the individual line profiles
corresponding to each radius $r$ produce at $V_{\rm obs}$.  The flux
in each ring, $2\pi\, r\, \Delta r\, \Sigma(r)$, thus serves as the
weighting factor for $\psi$ of each ring. We can now calculate the
weighted mean radius at each $V_{\rm obs}$ and identify for each value
of $V_{\rm obs}$ the radius which, in terms of weighted mean average,
is the dominant contributor to the total flux at that velocity. This
then makes it possible to determine the characteristic radii that give
rise to the flux at the velocities that define $W_{50}$.

Defining the weights $w_r = 2 \pi r\,\Delta r\, \Sigma(r)\,
\psi(V_{\rm obs}, V_c(r))$, the weighted mean radius $\bar{r}$ for each value of
$V_{\rm obs}$ is given by
\begin{equation}
\bar{r}(V_{\rm obs}) = \frac{ \sum_{r=0}^{R} w_r\, r}{\sum_{r=0}^{R} w_r}
\end{equation}

\begin{figure*}
\includegraphics[width=\hsize]{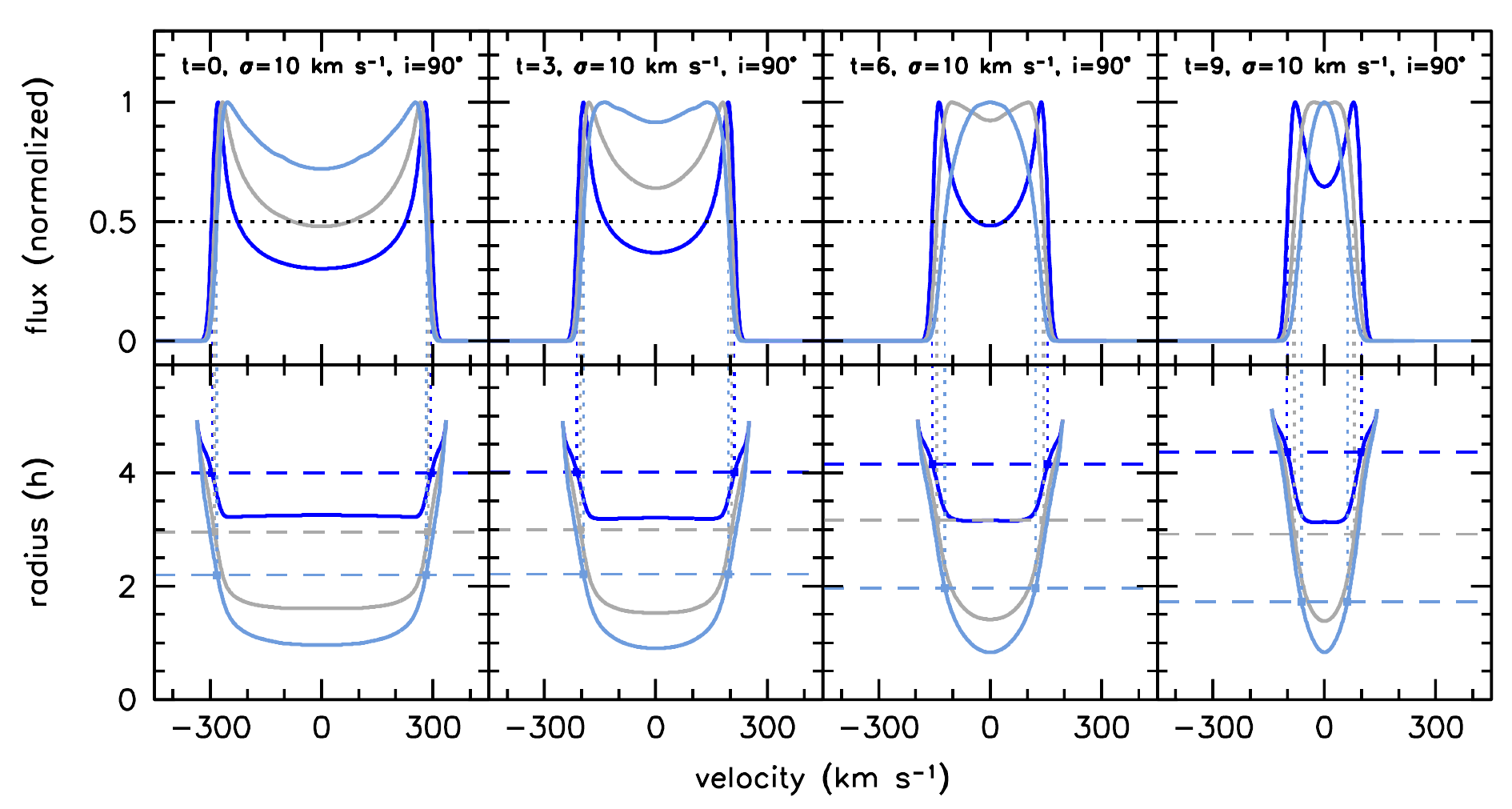}
\caption{The top row shows global profiles for $t = (0,3,6,9)$ as also
  shown in Fig.\ \ref{fig:profiles_i90}. Here only the flat (dark
  blue), exponential (grey) and S11 (light blue) profiles are
  shown. The bottom rown shows the mean weighted radius
  $\bar{r}(V_{\rm obs})$, as defined by Eqn.\ 4, that indicates which
  radius is the dominant contributor to the flux at a given value of
  $V_{\rm obs}$. Colors as in the top row. The dashed lines indicate
  the respective radii that dominate the flux at the velocities where $W_{50}$
  occurs. Vertical dotted lines connect the velocities at which $W_{50}$ occurs (in the top panels) with the corresponding radii (in the bottom panels).\label{fig:profradii}}
\end{figure*}

In Fig.\ \ref{fig:profradii} we plot the radii $\bar{r}(V_{\rm obs})$
for the $\sigma=10$ km s$^{-1}$ profiles that are also shown in
Fig.\ \ref{fig:profiles_i90}.  We also indicate the radii at which
$W_{50}$ occurs. We see a similar trend as the one in
Fig.\ \ref{fig:curvescale}, where the more compact density profiles
have smaller characteristic radii.  We can, analogous to Eqn.\ 4,
derive the weighted standard deviation of the radii contributing to
$V_{\rm obs}$ using
\begin{equation}
s(V_{\rm obs}) = \sqrt{ \frac{ \sum_{r=0}^{R} w_r \left(r - \bar{r}(V_{\rm
      obs})\right)^2}{\frac{N-1}{N} \sum_{r=0}^{R} w_r}}
\end{equation}
where $\bar{r}$ is the weighted mean from Eqn.\ 4, and $N$ is the
number of profiles of individual rings (radii) (as defined in Eqn.~1)
that have a non-zero contribution to the total global profile at
$V_{\rm obs}$ (as defined in Eqn.~2). For our models, $N$ will be large
and $(N-1)/N \approx 1$.

\begin{figure*}
\includegraphics[width=\hsize]{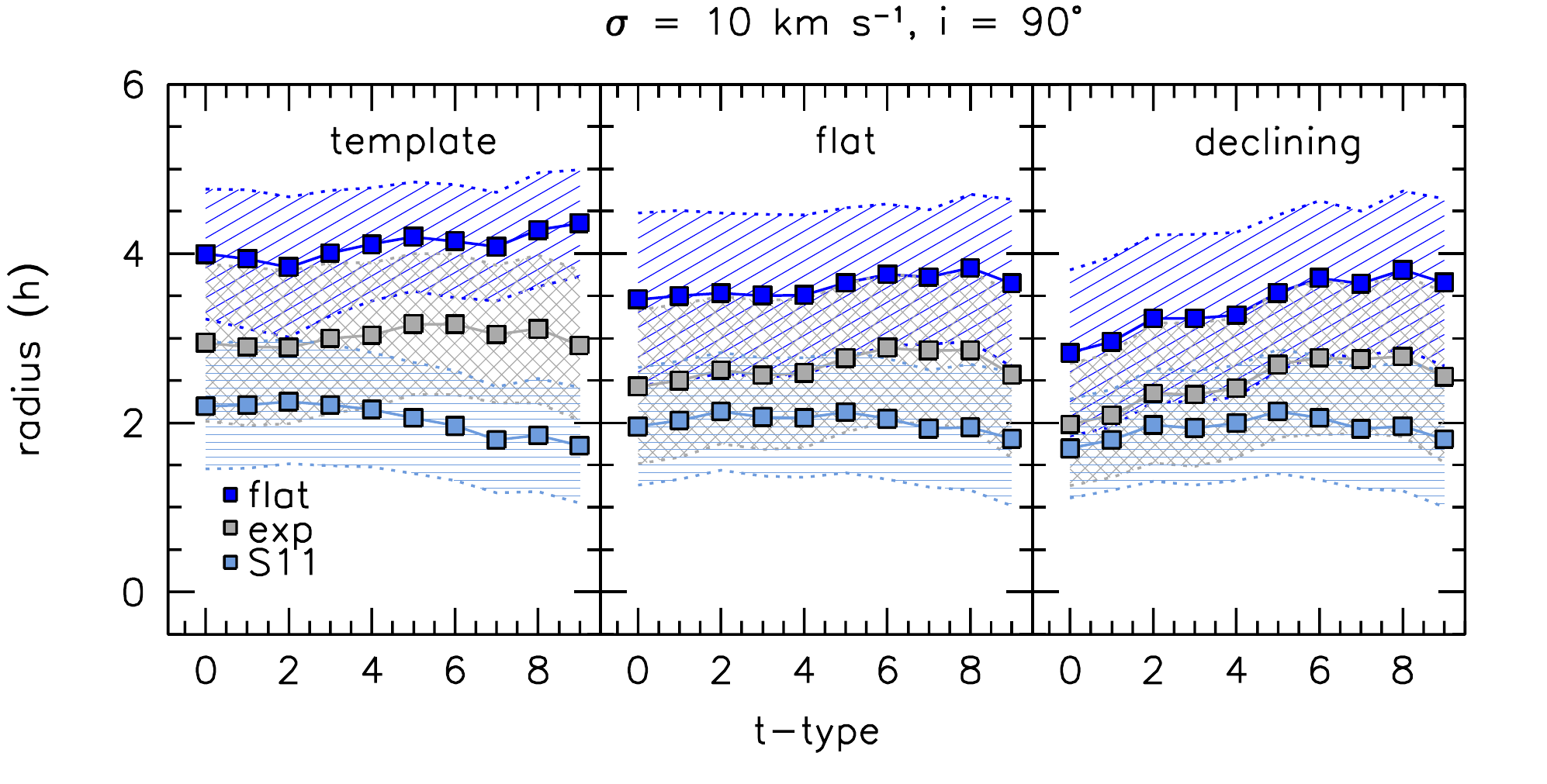}
\includegraphics[width=\hsize]{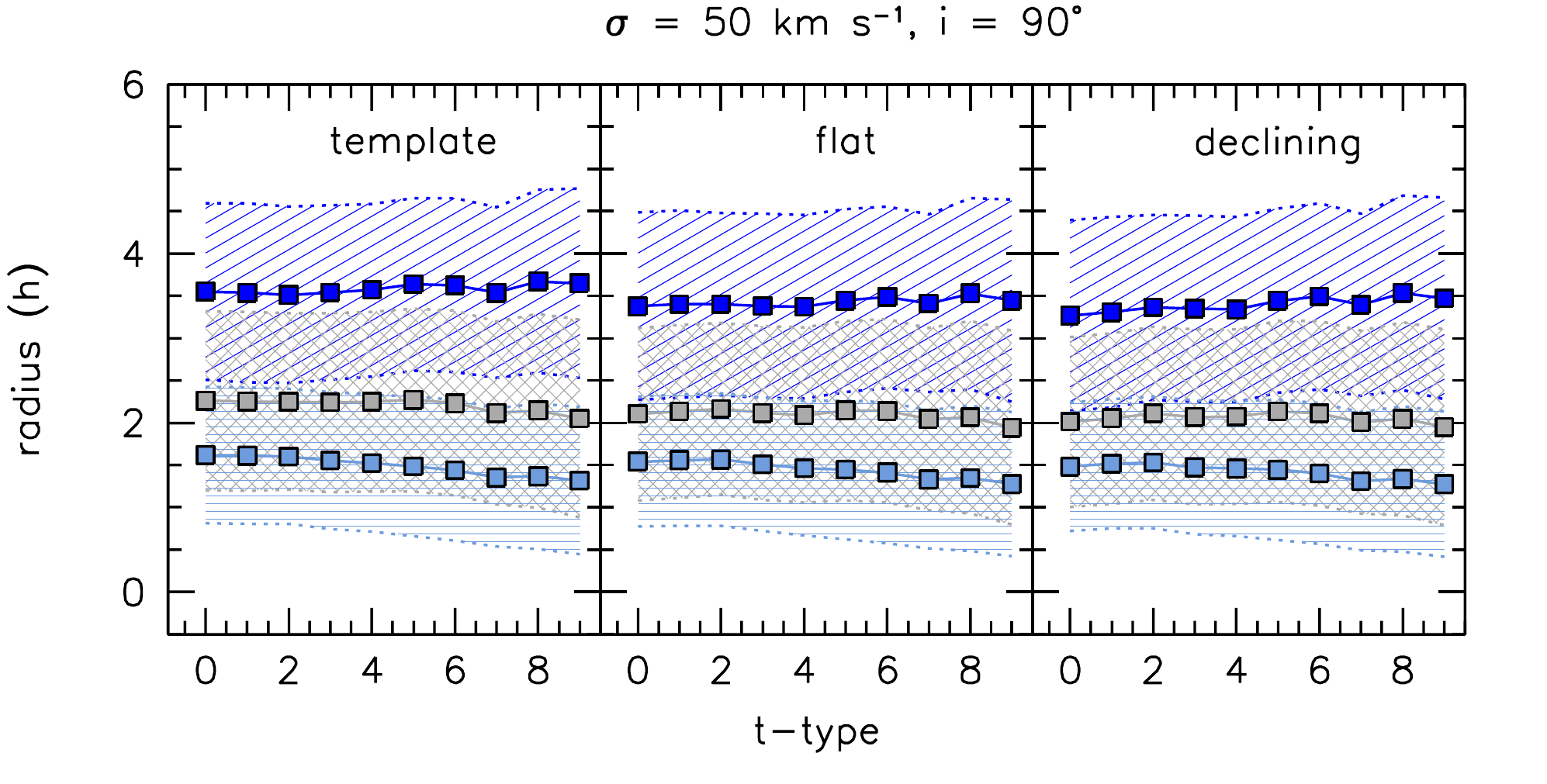}
\caption{Top row: values of the characteristic radius at which
  $W_{50}$ occurs (as defined in Eqn.~4) plotted as a function of
  $t$-type and assuming $\sigma = 10$ km s$^{-1}$. Shown are results
  for the flat density distribution model (dark blue), the exponential
  model (gray) and the S11 model (light blue). Hatched areas indicate
  the standard deviation in these radii as defined in Eqn.~5. The left
  panel shows the results for the \citet{cat06} models used in the
  majority of this paper; the center and right panels shows the
  results for the respective models with a flat and a declining
  rotation curve, as discussed in the text. Bottom row: same, but now
  for $\sigma=50$ km s$^{-1}$.
\label{fig:radii10}}
\end{figure*}

In Fig.\ \ref{fig:radii10} (top-left panel) we plot the mean radii
$\bar{r}$ where $W_{50}$ occurs (i.e., the equivalents of the radii
indicated in Fig.\ \ref{fig:curvescale}) as a function of $t$-type
assuming $\sigma = 10$ km s$^{-1}$. We also show the corresponding
standard deviation $s$, but again only present results for the flat,
exponential and S11 density profiles. The characteristic radii for
each density profile are well separated for all $t$-values.  The
average values for the radii we find are $(4.1 \pm 0.2)h$ for the flat
profile, $(3.0 \pm 0.1)h$ for the exponential profile, and $(2.0 \pm
0.2)h$ for the S11 profile (not taking into account the standard
deviations; the quoted uncertainties are the uncertainties in the
mean).  These radii agree very well with those derived based on the
rotation curves (see Sect.\ 4.1). Results for $\sigma = 50$ km
s$^{-1}$ are similar (Fig.\ \ref{fig:radii10}, bottom-left panel). 

\subsection{Dependence on rotation curve shape}

We test how robust the characteristic radii are against changes in the
shape of rotation curve.  One would expect that, to first order, the
velocity width derived from a global profile does not depend on the
precise shape of the rotation curve, but is determined mainly by the
amplitude of the rotation, the steepness of the density profile and
the velocity dispersion.  We recalculate global profiles for the
edge-on case, using the usual three density profiles, but now
assume flat or declining rotation curves. To do this we modify the
\cite{cat06} template polyex curves by changing the $\alpha$ parameter
as defined in Eqn.\ (3) and given in Tab.~1. This has the effect of
changing the outer slope, but does not affect the inner rise of the
rotation curve. For the flat rotation curves we simply set the value
for the outer slope $\alpha$ to zero. 

Choosing a value for $\alpha$ for the declining curves is more
complicated, as we want to be guided by the declining curves observed
in real galaxies, as opposed to modeling arbitrarily steeply declining
slopes. We therefore fit the polyex model to the two most steeply
declining H\,{\sc i}{} rotation curves in the THINGS sample
\citep{deblok08}, namely NGC 4736 and NGC 2903. Unfortunately, for
both these galaxies the inner slope is not tightly constrained (for
NGC 4736 due to absence of H\,{\sc i}{} in the center; for NGC 2903
due to streaming motions induced by the central bar), leading to some
degeneracy between the $r_{PE}$ and $\alpha$ fit parameters. For NGC
4736 we find values of $\alpha$ between $\sim -0.001$ and $\sim
-0.002$. When the part of the rotation curve of NGC 2903 that is
affected by the bar is not taken into account, we find $\alpha \sim
-0.002$. Fitting the full curve yields $\alpha \sim -0.018$, though
this value is likely affected by the uncertain inner rotation curve
and the associated uncertainy in $r_{PE}$.  Nevertheless, to be
conservative, we adopt a value $\alpha = -0.010$, which we regard as a
hard limit on the value of $\alpha$ found in real disk galaxies with
declining rotation curves. The results are shown in
Fig.\ \ref{fig:radii10}.

We see that the characteristic radius of the flat density distribution
has decreased somewhat for both flat and declining rotation curves,
but they do not differ dramatically from the ones found for the
original template rotation curves.  Given the extreme outer slope we have
adopted for the declining curve, it is therefore still justified to
use a characteristic radius of $\sim 4h$ to $\sim 5h$ for determining
the dynamical mass using any global profile that assumes a flat tracer
profile.

The characteristic radii for the S11 distribution also show little
change and remain around $2h$.  The results for $\sigma = 50$ km
s$^{-1}$ show much less difference when comparing the template, flat
and declining rotation curve models, due to the more dominant velocity
dispersion. Nevertheless, to first order the radii derived for the
low-dispersion case also hold here.  These results thus show that
  the shape of the global profile is not critically dependent on the
  shape of the rotation curve.

\section{Summary}

The global velocity profile of a galaxy is often used as a means to
estimate its dynamical mass.  The profile is determined by the
rotation curve of a galaxy, the gas distribution within a galaxy and
the velocity dispersion of the gas. We have investigated how the
global profile changes as a function of these parameters. This is
important due to the different ways global velocity profiles are
typically determined at low and high redshifts.  Local measurements of
global velocity profiles mostly use H\,{\sc i}{} observations.  Local
H\,{\sc i}{} disks usually have modest velocity dispersions, are
extended with an almost constant surface density and sample the outer
parts of the rotation curves well. On the other hand, CO (the main ISM
tracer at high redshift) has a radial distribution that (at least for
nearby galaxies) is known to be much steeper with an exponential
decline. In addition, the velocity dispersion at high redshift is
probably higher than in local galaxies. The difference in properties
and distribution of the tracers can thus lead to differences in global
profile widths and  dynamical masses.

While the results presented here are generally applicable, for a
comparison with real high-redshift galaxies we have to assume 1) that
these galaxies are virialised and dominated by rotation; 2) that they
are not majorly affected by interactions and/or asymmetries (though
the latter will only have a small impact on $W_{50})$; 3) that
rotation curves at high-redshift are to first order the same as at low
redshift; 4) and that the range of gas density profiles modeled here
overlaps with that in real galaxies.  Choosing a different gas density
profile from the range modelled here will account for the possibility
that in high-redshift galaxies the gas distribution may be different
from that in a low-redshift galaxy.

A first conclusion from the modeling of these profiles is that higher
velocity dispersion and steeper density profiles result in less
pronounced double-horned profiles. This is consistent with
high-redshift CO observations of galaxies, as discussed in Sect.~3.1. 
Galaxies with identical dynamics will  have different global
profiles, depending on the gas tracer used in the observations.

The steeper density profiles associated with CO give smaller velocity
widths for otherwise identical rotation curves.  The ratio between CO
and H\,{\sc i}{} $W_{50}$ values changes from $\sim 0.95$ for the most
luminous galaxies we modeled ($t=0$, or $M_I \sim -23.8$) to $\sim
0.6$ for the faintest galaxies ($t=9$ or $M_I \sim -19.0$).

This change affects the inferred slope of the TF-relation as well. The
template rotation curves we use, in combination with a flat,
constant-density distribution of the gas (`H\,{\sc i}{}'), give a $W_{50}$-based
TF slope of $\sim 3.7$. Using an S11 distribution for the gas (`CO')
instead, decreases the slope to $\sim 2.6$. These numbers assume a 10
km s$^{-1}$ velocity dispersion, but the same decrease in slope is
observed for the 50 km s$^{-1}$ models.

When global profiles are used for dynamical mass estimates, resolved
observations are usually not available and a radius must be assumed to
calculate the dynamical mass. Usually an optical radius is used for
this, though the definition of this radius not well constrained. To get
around having to make an arbitrary choice for a radius, we calculate
the effective radius of the rotation curve that gives rise to the
measured $W_{50}$ value for a number of combinations of rotation
curves and density profiles. We find that each gas density profile
results in its own narrow range of characteristic radii, thus enabling
a determination of the dynamical mass using a velocity width and a
radius that is appropriate for the tracer used. Velocity widths
$W_{50}$ measured in H\,{\sc i}{} (with a flat density profile)
typically originate at $\sim 5h$,  while velocity widths derived
  using the S11 density profile (equivalent to low-redshift CO
  distributions) originate at $\sim 2h$. These results thus give us,
for a range in gas density profile slopes, a consistent definition for
the length scales to use in determining dynamical masses.

\acknowledgements
We thank Barbara Catinella for making available electronic versions of
the template rotation curves.  WJGdB was supported by the European
Commission (grant FP7-PEOPLE-2012-CIG \#333939).

\clearpage

\appendix

\section{Comparison with three-dimensional models}

The analytical method described in \citet{obr09} is an efficient way
to calculate a large number of global profiles. Here we compare some
of these analytical global profiles with profiles derived from full
three-dimensional models of rotating galaxies. Constructing the latter
models is significantly more CPU-intensive, but it enables us to
derive global profiles using methods that are also used when analysing
real data.  Good agreement between the two sets of profiles thus
ensures that the conclusions we derive from the analytical models also
apply to global profiles of real galaxies.

To construct the data cube models we use the TiRiFiC package
\citep{tir07}. TiRiFiC distributes cloud particles through a
user-created three-dimensional (right ascension, declination,
velocity) data cube, using user-defined geometric and kinematic
parameters as distribution functions. Among the parameters that can be
defined are the rotation curve, radially varying inclination and
position angles, the surface density profile, the vertical
distribution and extent of the gas and the velocity
dispersion. TiRiFiC produces a data cube that can then be ``observed''
and treated like a real observation, and global profiles can be
constructed using standard data analysis methods.

\begin{figure*}
\includegraphics[width=0.9\hsize]{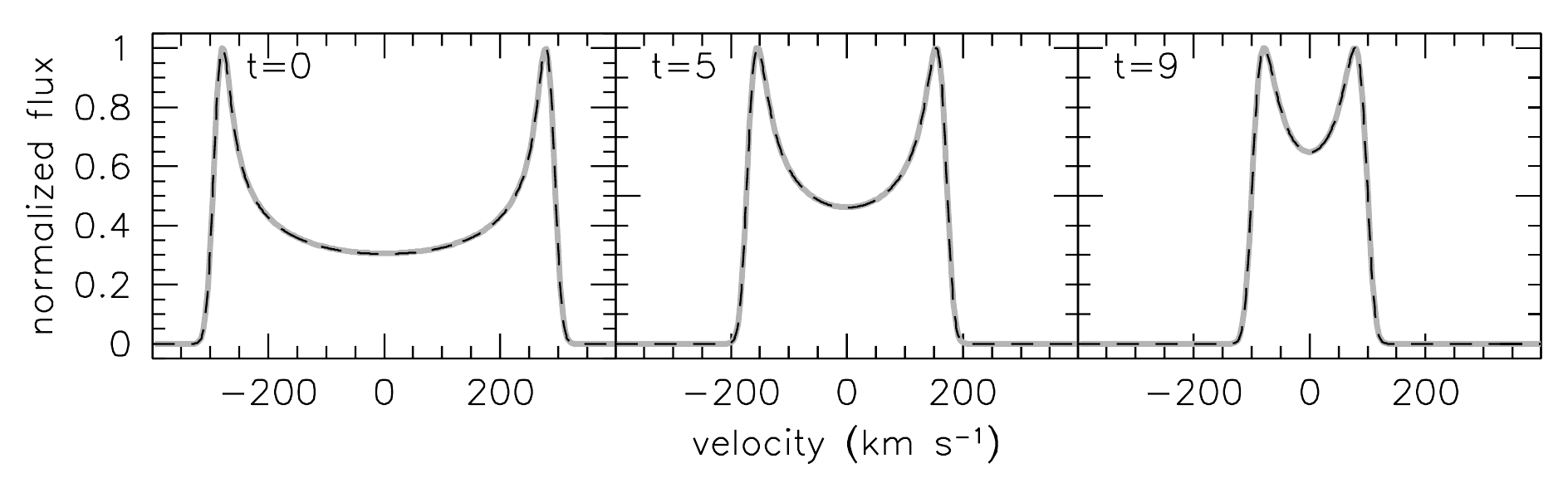}
\caption{Comparison of global profiles: Grey full profiles are from full
  three--dimensional TiRiFiC models, the black dashed profile is from
  the analytical description. $t$ values refer to template curves. As
  the curves are practically indistinguishable we use the
  analytic curves as they are computationally much less expensive.
\label{fig:comparemods}}
\end{figure*}

We have compared global profiles produced by both methods for a large
range of template rotation curves, gas distributions, inclinations and
dispersions, and find the differences between the two methods to be
negligible.  As an example, we compare in Fig.~\ref{fig:comparemods}
the global profiles of three different edge-on galaxies, each with a
different template rotation curve. A constant radial gas surface
density was assumed for all three examples.  We see very good
agreement between the profiles from our three-dimensional models and
the analytic description.  For the profiles shown in
Fig.~\ref{fig:comparemods}, the mean absolute difference is 0.15 per
cent; the maximum absolute difference is 0.4 per cent. These
differences are purely due to specific choices made in the
construction of the data cubes (e.g., pixel size and channel spacing)
of the TiRiFiC models and are not systematic.  For all practical
purposes the profiles are identical and we proceed to use the
analytical profiles derived using the \citet{obr09} method.

\section{Other inclinations}

In the main text we have only discussed the case of edge-on
galaxies. For these galaxies the signal of rotation is maximal and the
effects of dispersion minimised. Here we explore how that balance
changes for lower inclinations, and whether there is a significant
impact on the measurements of velocity widths.

We investigate two additional cases: one where the inclination is
60$^{\circ}$, the average inclination of a randomly oriented sample of
galaxies, and one where the inclination is 30$^{\circ}$, a practical
lower limit on the inclination of real galaxies where reliable
measurements of the rotation can still be done. For even lower
inclinations the uncertainties in the associated inclination
corrections become too large.

\begin{figure*}
\includegraphics[width=0.9\hsize]{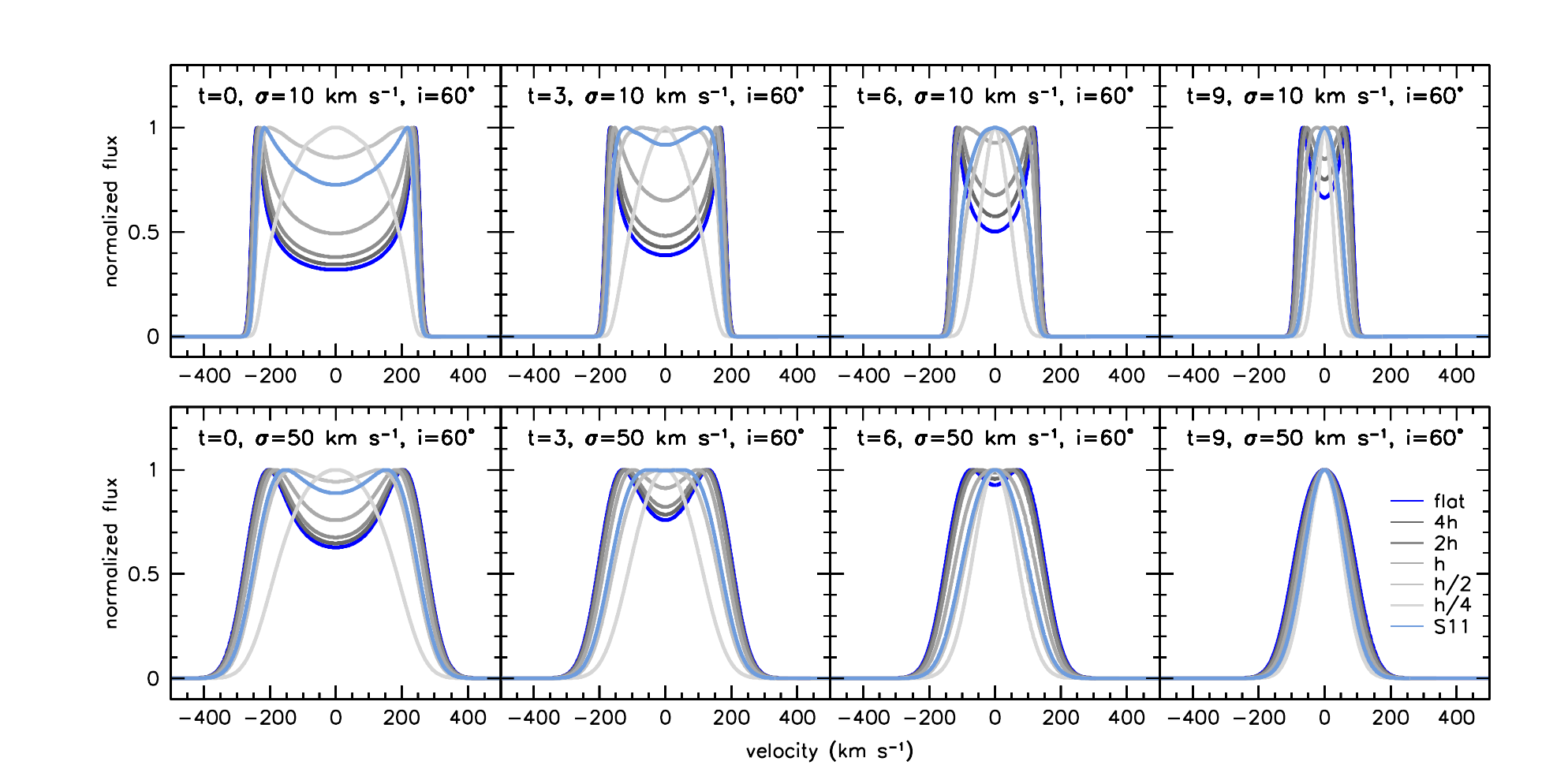}
\caption{As Fig.\ \ref{fig:profiles_i90}, but assuming $i=60^{\circ}$.
\label{fig:profiles_i60}}
\end{figure*}
\begin{figure*}
\includegraphics[width=0.9\hsize]{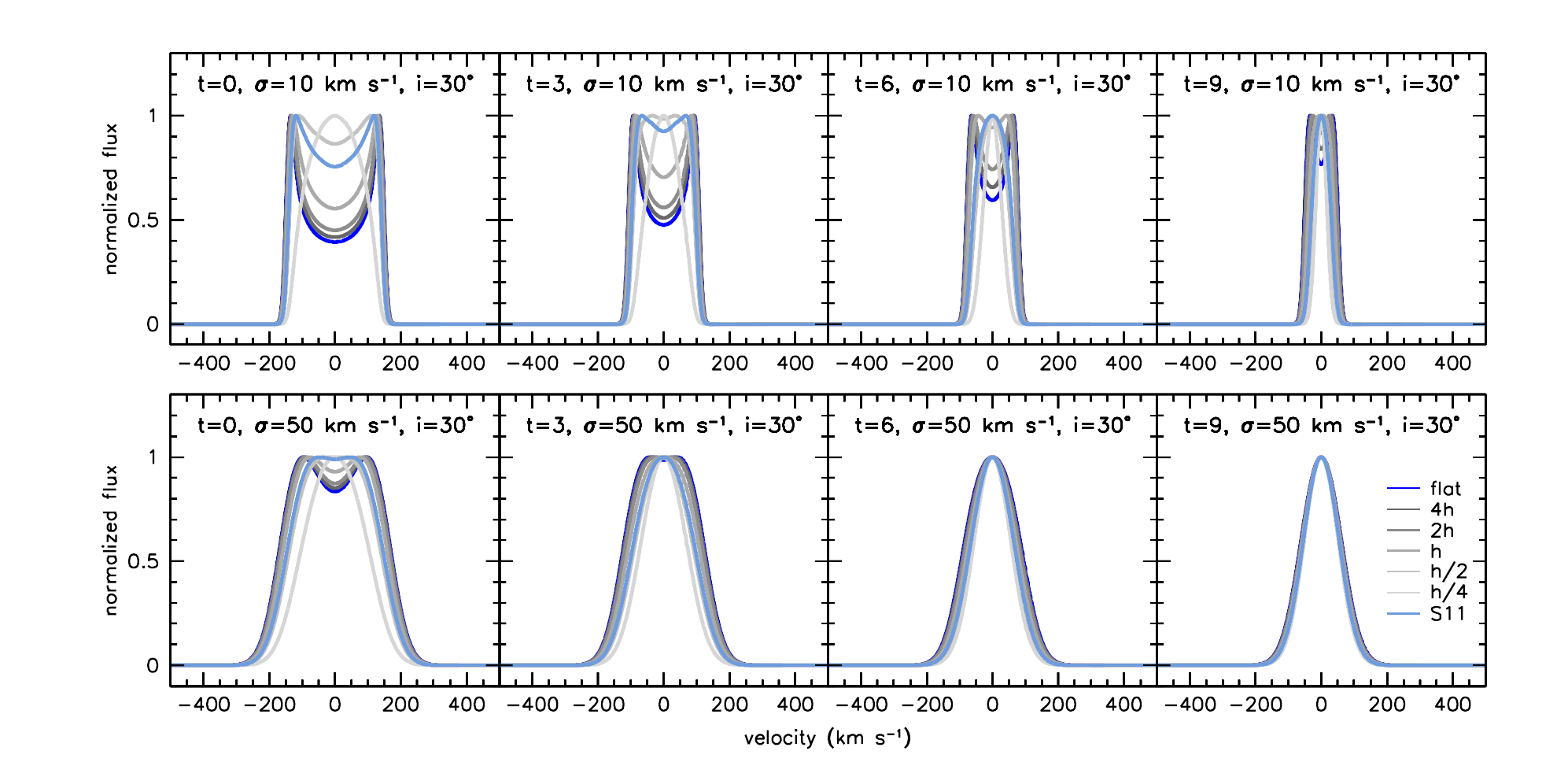}
\caption{As Fig.\ \ref{fig:profiles_i90}, but assuming $i=30^{\circ}$.
\label{fig:profiles_i30}}
\end{figure*}

\begin{figure*}
\includegraphics[width=0.9\hsize]{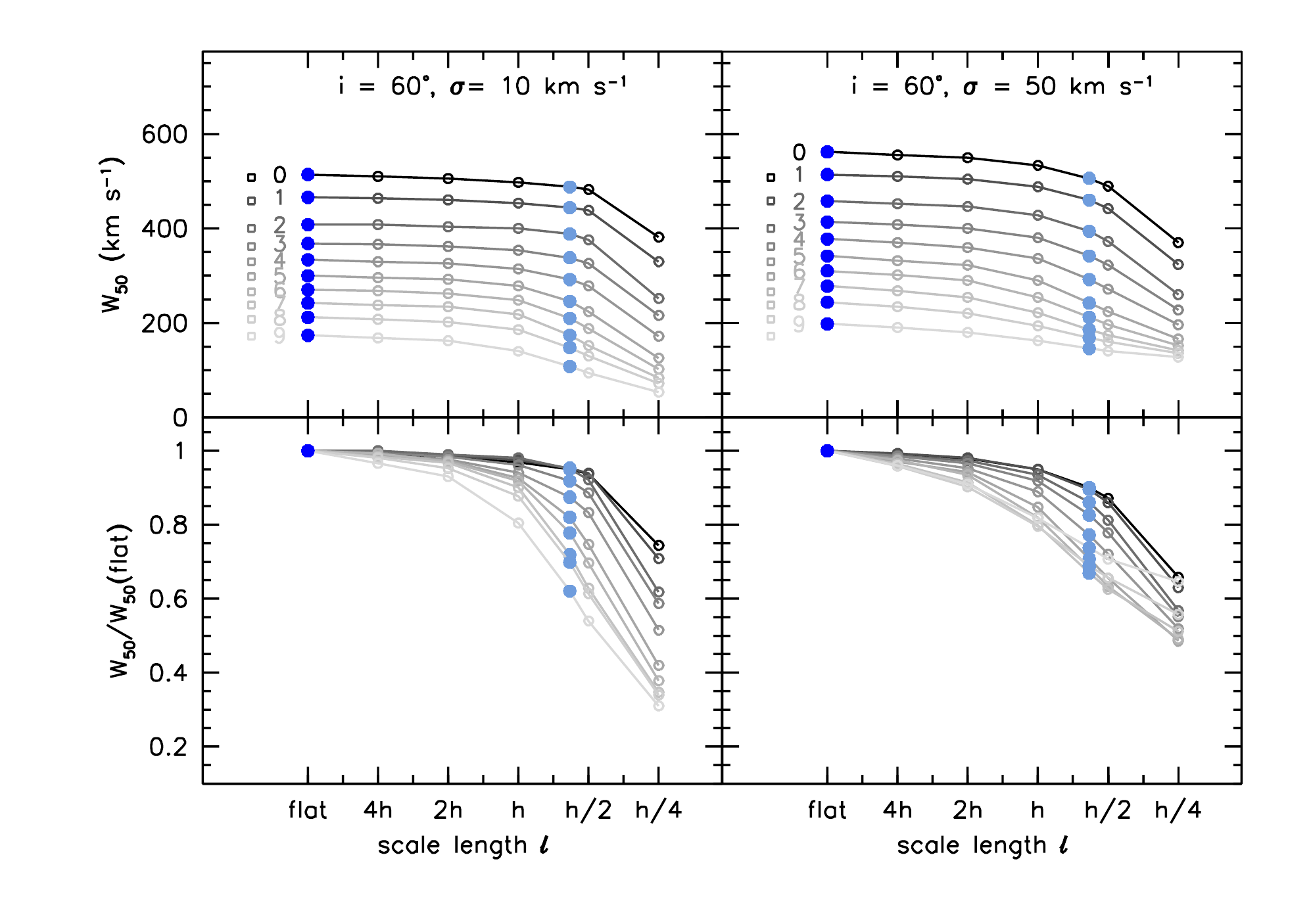}
\caption{As Fig.\ \ref{fig:w50}, but assuming
  $i=60^{\circ}$.\label{fig:w50_i60}}
\end{figure*}

\begin{figure*}
\includegraphics[width=0.9\hsize]{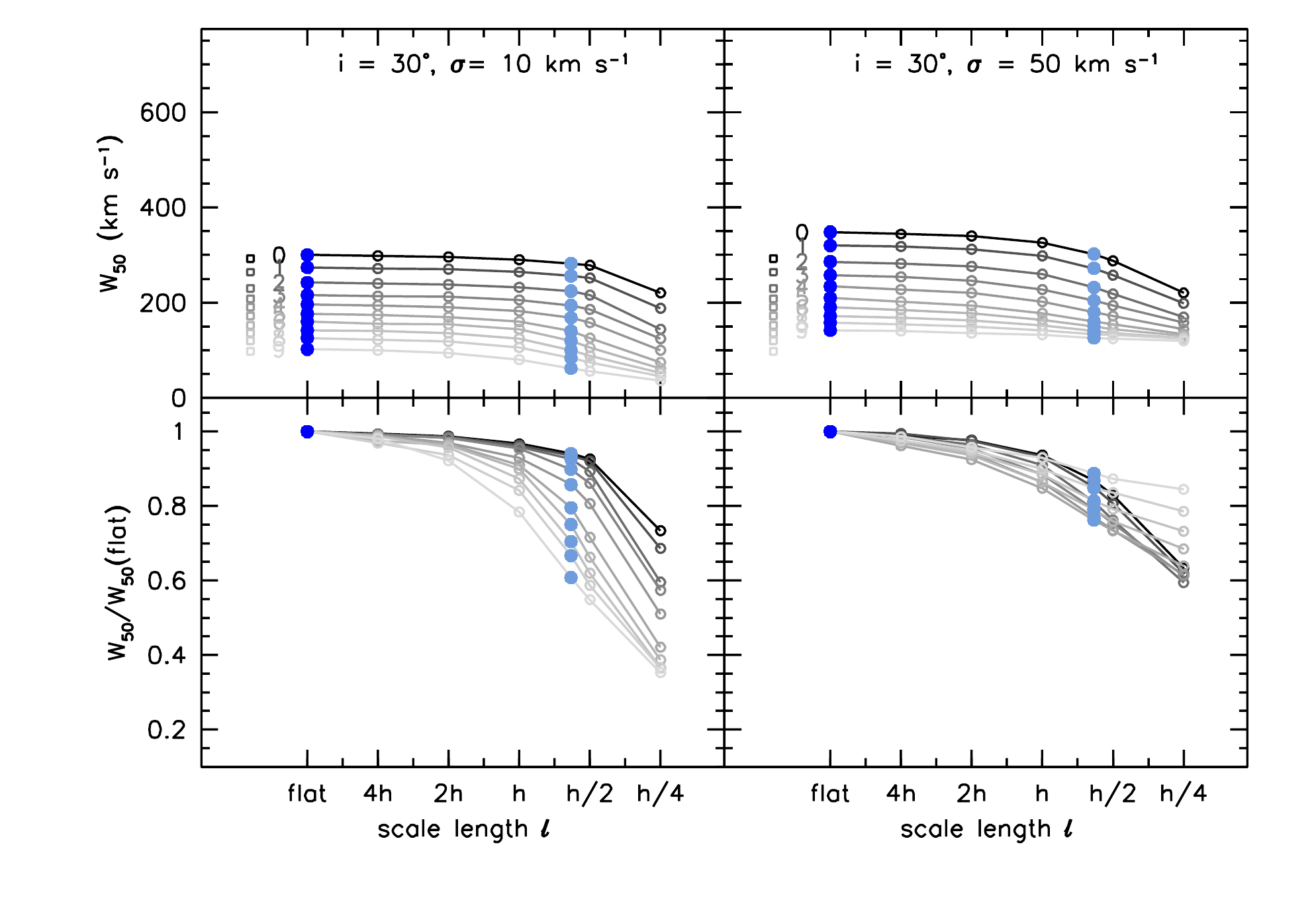}
\caption{As Fig.\ \ref{fig:w50}, but assuming
  $i=30^{\circ}$.\label{fig:w50_i30}}
\end{figure*}

Figures \ref{fig:profiles_i60} and \ref{fig:profiles_i30} show the
global profiles for the same galaxies as shown in Fig.~1, but now
assuming $i=60^{\circ}$ and $i=30^{\circ}$, respectively. It is clear
that the effect of dispersion and profile steepness has become more
pronounced. Especially for $i=30^{\circ}$ and $\sigma=50$ km s$^{-1}$
(bottom-row in Fig.\ \ref{fig:profiles_i30}) we see that there is hardly any
signature of rotation left, except for the most massive galaxies
(lowest $t$-types). This is also seen in the changes in $W_{50}$ as a
function of profile steepness as shown in Figs.\ \ref{fig:w50_i60} and
\ref{fig:w50_i30}. 

For $\sigma = 10$ km s$^{-1}$ and $i=60^{\circ}$ there are only minor
changes. Profile widths have shifted downwards due to the inclination,
but in the steep density slopes part of the diagram they have more or
less converged to the values found for the edge-on case, reflecting
the very limited radial range that is probed in the latter case.
Similar conclusions can be drawn for the $\sigma = 50$ km s$^{-1}$
case. 

For the $i=30^{\circ}$ case we see that for $\sigma = 10$ km s$^{-1}$,
apart from a much compressed range in $W_{50}$, the trends in velocity
width with density profile steepness are similar as for $i=90^{\circ}$
and $i=60^{\circ}$. This situation is very different for the 50 km
s$^{-1}$ velocity dispersion. Here the large amplitude of the
dispersion compared to the projected rotation velocity causes the
velocity widths to be dominated by dispersion, especially for the
steep density profiles and/or high $t$-type galaxies. This also means
that the corresponding derived dynamical masses are not representative
of the intrinsic masses.

\begin{figure*}
\centerline{\includegraphics[width=0.4\hsize]{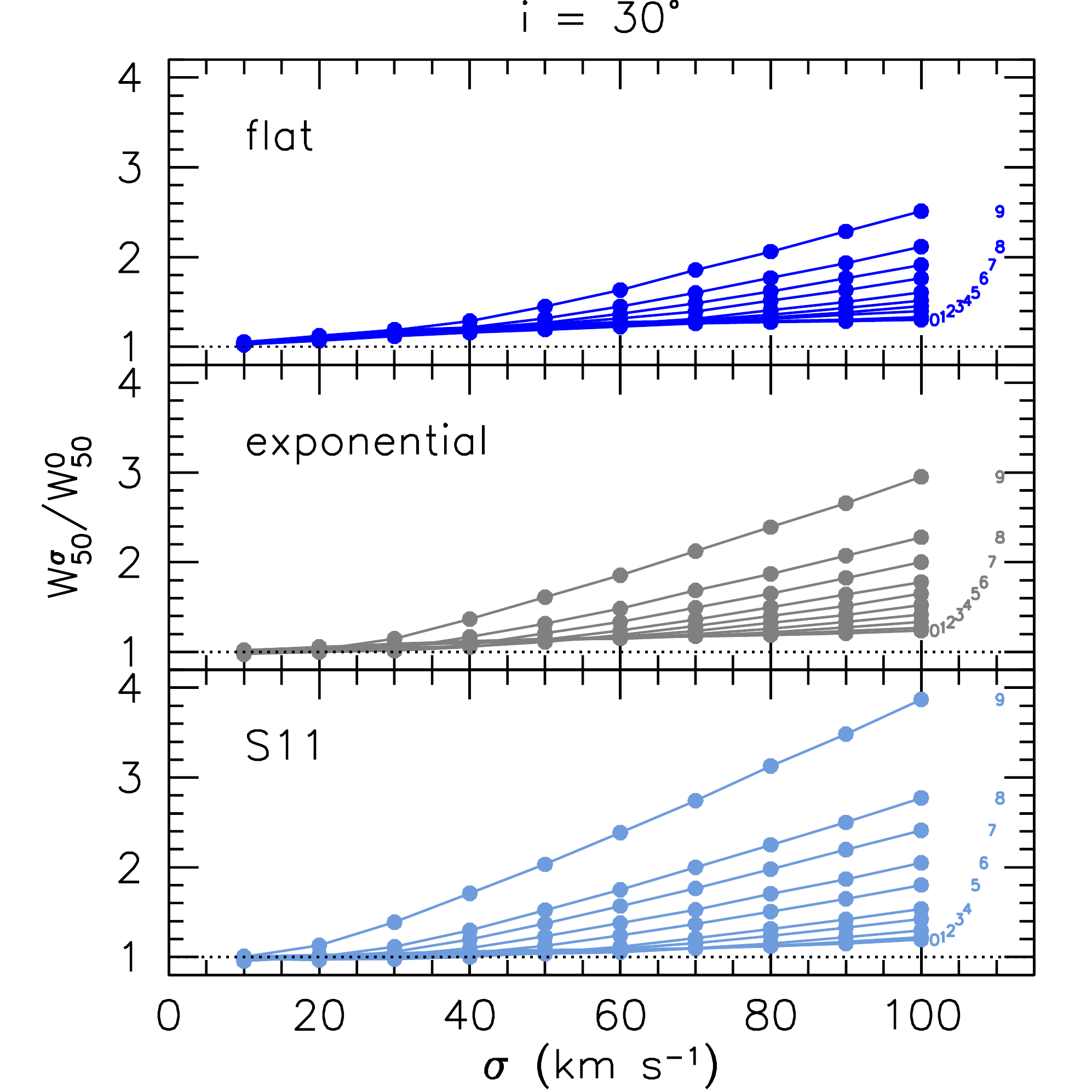}
\includegraphics[width=0.4\hsize]{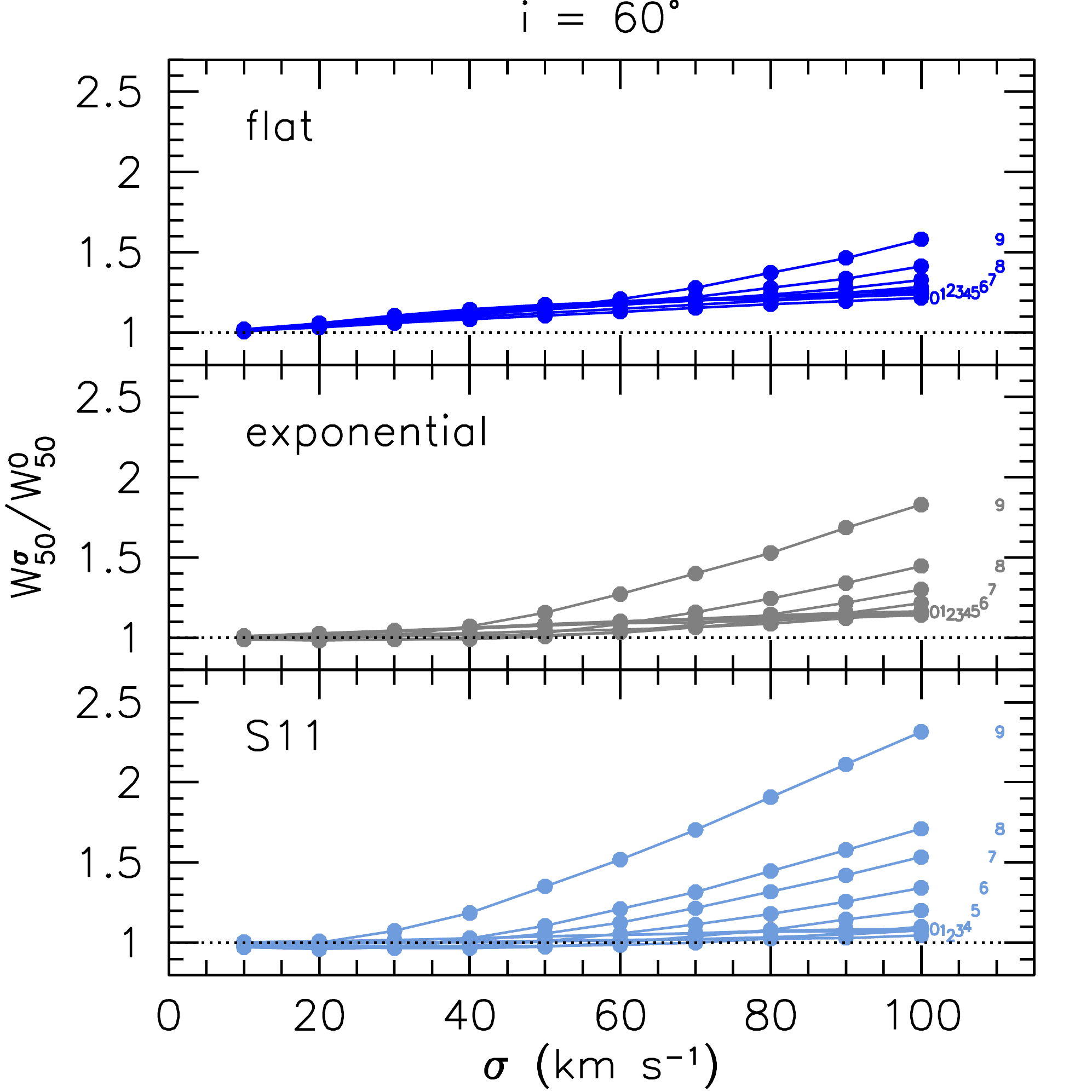}}
\caption{Ratio of the velocity widths that include velocity dispersion
  (as indicated on the horizontal axis) and the zero-dispersion
  velocity widths. The left panel shows corrections for $i=30^{\circ}$,
  the right panel for $i=60^{\circ}$. In each of the panels, from top
  to bottom are shown the ratios for the flat, exponential and S11
  density profiles. The correction curves are shown for each $t$-type,
  as indicated by the numbers in the right-most part of the plot.
\label{fig:dispratio30_60}}
\end{figure*}

Lastly we investigate the ratio of the velocity widths of profiles
including a dispersion term and those with zero dispersion, as shown
for edge-on galaxies in Fig.~\ref{fig:dispratio}, but here evaluated
for $i=60^{\circ}$ and $i=30^{\circ}$. These are shown in
Fig.\ \ref{fig:dispratio30_60}. It is clear that corrections become
more dominant for lower inclination galaxies.

\end{document}